\documentclass[%
 reprint,
nofootinbib,
longbibliography,
noeprint,
 amsmath,amssymb,
 aps,
]{revtex4-2}

\usepackage{comment}
\usepackage{cases}
\usepackage{graphicx}
\usepackage{dcolumn}
\usepackage{bm}
\usepackage{siunitx}
\usepackage{xcolor}
\usepackage{hyperref}
\usepackage{scrextend}
\definecolor{darkblue}{rgb}{0.0,0.0,0.3}
\definecolor{nred}{rgb}{0.7,0.3,0.3}
\hypersetup{colorlinks,breaklinks,
            linkcolor=darkblue,urlcolor=darkblue,
            anchorcolor=darkblue,citecolor=darkblue}
\usepackage[]{cleveref}

\usepackage{ulem}
\crefname{section}{\S\!}{\S\S\!}
\crefname{appendix}{App.}{Apps.}
\crefname{equation}{Eq.}{Eqs.}
\Crefname{equation}{Equation}{Equations}
\crefname{figure}{Fig.}{Figs.}
\Crefname{figure}{Figure}{Figures}

\newcommand{\bzpm}{\tilde{\bm{z}}^{\pm}}
\newcommand{\bzmp}{\tilde{\bm{z}}^{\mp}}
\newcommand{\bzm}{\tilde{\bm{z}}^{-}}
\newcommand{\bzp}{\tilde{\bm{z}}^{+}}
\newcommand{\zm}{\tilde{z}^{-}}
\newcommand{\zp}{\tilde{z}^{+}}
\newcommand{\zpm}{\tilde{z}^{\pm}}
\newcommand{\zmp}{\tilde{z}^{\mp}}
\newcommand{\egrad}{\hat{\nabla}} 
\newcommand{\egradp}{\hat{\nabla}_{\perp}} 
\newcommand{\gradp}{\tilde{\nabla}_\perp} 
\newcommand{\bb}{\tilde{\bm{b}}_\perp}
\newcommand{\bu}{\tilde{\bm{u}}_\perp}
\newcommand{\va}{v_{\rm A}}
\newcommand{\omegaa}{\omega_{\rm A}}
\newcommand{\vao}{v_{\rm A0}}
\newcommand{\zh}{\hat{\mathbf{z}}}
\newcommand{\xh}{\hat{\mathbf{x}}}
\newcommand{\yh}{\hat{\mathbf{y}}}
\newcommand{\zetap}{\tilde{\zeta}^{+}}
\newcommand{\zetam}{\tilde{\zeta}^{-}}
\newcommand{\zetapm}{\tilde{\zeta}^{\pm}}
\newcommand{\zetamp}{\tilde{\zeta}^{\mp}}
\newcommand{\az}{\tilde{A}_z}
\newcommand{\phit}{\tilde{\Phi}}
\newcommand{\anas}{\tilde{\mathcal{A}}}
\newcommand{\Lp}{\tilde{L}_{+}}
\newcommand{\lamp}{\tilde{\lambda}_{+}}
\newcommand{\lamm}{\tilde{\lambda}_{-}}
\newcommand{\taup}{\tau_{+}}
\newcommand{\taum}{\tau_{-}}
\newcommand{\lampm}{\tilde{\lambda}^{\pm}}
\newcommand{\mB}{\overline{\bm{B}}}
\newcommand{\rmd}{{\rm d}}
\newcommand{\ffold}{figures}

\begin{document}

\title{Reflection-driven turbulence in the super-Alfv\'enic solar wind}

\author{Romain Meyrand}
 \affiliation{Department of Physics, University of Otago, 730 Cumberland St., Dunedin 9016, New Zealand}
\author{Jonathan Squire}%
\affiliation{Department of Physics, University of Otago, 730 Cumberland St., Dunedin 9016, New Zealand
}%
\author{Alfred Mallet}
\affiliation{Space Sciences Laboratory, University of California, Berkeley CA 
94720, USA
}%
\author{Benjamin D. G. Chandran}
\affiliation{Department of Physics and Astronomy, University of New Hampshire, Durham, NH 03824, USA
}


\date{\today}

\begin{abstract}
In magnetized, highly stratified astrophysical environments such as the Sun's corona and solar wind,   Alfv\'enic fluctuations  ``reflect'' from background gradients, enabling nonlinear interactions and thus  dissipation of their energy
into heat. This process, termed ``reflection-driven turbulence,'' is thought to play a crucial role in coronal heating and solar-wind acceleration, 
explaining a range of detailed observational correlations and constraints.
Building on previous works focused on the inner heliosphere, 
here we study the basic physics of reflection-driven turbulence using reduced magnetohydrodynamics in an expanding box---the 
simplest model that can capture the local turbulent plasma dynamics in the super-Alfv\'enic solar wind. Although  
idealized, our high-resolution simulations and simple theory reveal a rich phenomenology that is consistent with a diverse range of observations.  Outwards-propagating fluctuations, which initially have high imbalance (high cross helicity),  decay nonlinearly to heat the plasma, becoming more
balanced and magnetically dominated. 
Despite the high imbalance, the turbulence is strong because Els\"asser collisions are suppressed by reflection, leading to ``anomalous coherence'' between the two  Els\"asser fields.
This coherence, together with linear effects, causes the turbulence to anomalously grow the 
``anastrophy'' (squared magnetic potential) as it decays, forcing the energy 
to rush to larger scales and forming a ``$1/f$-range'' energy spectrum as it does so. At late times, 
the expansion overcomes the nonlinear and Alfv\'enic physics, 
forming isolated, magnetically dominated ``Alfv\'en vortex'' structures that minimize their nonlinear dissipation. 
These results can plausibly explain the observed radial and wind-speed dependence of turbulence imbalance (cross helicity), residual energy, plasma heating,
and fluctuation spectra, as well as making a variety of testable predictions for future observations. 
\end{abstract}

\maketitle

\section{Introduction}

The mechanisms that heat and accelerate the solar wind remain mysterious, or at least controversial \cite{Cranmer2019}. 
In order to explain decades of \textit{in-situ} spacecraft data, particularly local temperature measurements 
and the  high speeds of fast-wind streams, there must exist an energy source to heat the plasma even 
at large distances from the solar surface. A leading paradigm for explaining this extended heating is  {Alfv\'enic turbulence}, in which  
the energy is provided by Alfv\'en waves 
launched from  the low solar atmosphere by photospheric motions or magnetic reconnection \cite{axford92,depontieu07}.
 As these waves propagate outwards, away from the Sun, they become turbulent, causing their energy to cascade to smaller scales and dissipate \cite{Velli89,Dmitruk02,dmitruk03,cranmer05}. The resulting turbulent heating increases the plasma pressure, which, along with the wave pressure, accelerates the solar wind away from the Sun \cite{tu87,tu88,cranmer07,verdini10}.

Although plausible, particularly given the extended turbulent-like fluctuations observed in the solar-wind plasma \cite{Belcher1971,Bruno2013,Kiyani2015,Chen2016}, 
a particular difficulty with this model lies in the robustness of Alfv\'enic fluctuations: in a homogenous plasma, Alfv\'en waves propagating in the same direction do not interact with one another or damp out, even at 
large amplitudes and/or when their wavelength is well below the mean free path \cite{Barnes1974,Kulsrud1983}. Turbulence, as likely needed to dissipate their energy, 
thus arises only via interactions between the two ``Els\"asser'' fields $\bm{z}^\pm$, 
which are the  counter-propagating linear eigenmodes in a homogenous plasma \cite{Elsasser1950,iroshnikov63,kraichnan65}.
With the Sun supplying energy only in outwards-propagating waves  dominated by one Els\"asser field, some source of the other Els\"asser field is needed to generate turbulence that could explain the observed heating.
One possible mechanism for enabling this process is reflection arising from the radial variation in the background Alfv\'en speed~$\va$ \cite{heinemann80,velli93}. The turbulence that results due to this interaction between outwards and reflected waves is generally referred to as ``reflection-driven turbulence'' \cite{Velli89}.
Phenomenological models and simulations suggest that the paradigm can
broadly explain many observed local and global features of the solar wind \cite{cranmer05,verdini07,Chandran09a,Verdini12,Perez13,Ballegooijen17,Shoda2022},
although there remain important unresolved issues and questions \cite{Shoda2019,AsgariTarghi2021,Chandran2021}.
Similar mechanisms may also play a key role in  other astrophysical systems with large density gradients and strong magnetic fields, 
particularly compact-object accretion flows, which are known to possess hot, compact corona that are likely fed by strong fluctuations in the disk below \cite{Reis2013,Chandran2018}.

The goal of this work is to study reflection-driven turbulence from the most basic standpoint possible, elucidating 
the key features in a simplified setting. 
This differs from previous studies, which have usually used either phenomenological models \cite{Dmitruk02,verdini07,Chandran09a,Reville2020,Shoda2022}
or radially extended ``flux-tube'' simulations \cite{dmitruk03,Perez13,vanballegooijen16,Ballegooijen17,Chandran19} to attempt to realistically match observed parameters and regimes of the corona and solar wind. 
Both perspectives---the basic and the realistic---are important, but we believe the former has been 
neglected in previous literature. Rectifying this omission is especially relevant  because reflection-driven turbulence is neither decaying
nor forced (the two limits usually considered in turbulence studies), meaning that care is needed when applying intuitions and ideas from broader turbulence research.

 Our approach is to use the so-called ``expanding box model'' (EBM) \cite{Grappin93}, which 
 tracks a small parcel of plasma as it flows away from the Sun. The version of the EBM we use applies to regions beyond the Alfv\'en radius (or surface) $R_{\rm A}$ where the solar-wind speed $U$ overtakes the Alfv\'en speed and becomes approximately constant with radius $R$. This local approach also differs from most previous work on reflection-driven turbulence (although the EBM has proved important in other solar-wind and turbulence contexts \cite{Grappin96,Dong14,Montagud18,Squire2020,Grappin2022,Johnston22}). A  disadvantage of the EBM is that our results 
cannot be applied directly to the solar-wind acceleration region (although some aspects may prove translatable); an 
advantage is the simplicity of using a homogenous, periodic domain, which allows for much higher numerical resolutions and 
decreases the number of free parameters while capturing many of the essential physical ingredients. In 
addition,  our results seem to explain 
a variety of disparate observations from \textit{in-situ} spacecraft measurements at $R>R_{\rm A}$, 
some of which have been missed in previous theoretical works because of the focus on 
lower-altitude acceleration regions. We argue that these observational comparisons provide persuasive 
evidence that reflection-driven turbulence controls important aspects of solar-wind turbulent evolution 
beyond $R_{A}$, as well as providing a number of testable and falsifiable predictions for future works.

As well as the contributions described above, our main novel result is that 
reflection-driven turbulence precipitates a strong inverse energy transfer as it decays. 
This feature, which we argue is a consequence of an anomalous conservation law for the squared 
parallel magnetic vector potential (``anastrophy''), causes initially small-scale 
outwards-propagating fluctuations to rush to large scales as they decay, forming a ${\propto}k_{\perp}^{-1}$ spectrum 
in the process (here $k_{\perp}$ is the wavenumber perpendicular to the background magnetic field). 
This suggests the observed large-scale fluctuations that dominate the solar-wind turbulence spectrum can develop \textit{in situ} as the
wind propagates, which may be important if low-frequency waves 
are unable to effectively propagate through the chromosphere-coronal transition due to 
large local  gradients in the Alfv\'en speed \cite{Leroy1981,velli93,Ballegooijen17,Reville18}.
Another new result concerns the asymptotic evolution of the
turbulence at large radii, where it becomes governed by large-scale magnetically dominated ``Alfv\'en vortices'' \cite{Petviashvili92,Alexandrova08}. 
These structures, which are approximate nonlinear solutions and so dissipate into heat only very slowly, tend to freeze into the plasma
at late times, growing continuously as the plasma expands. 

The remainder of the paper is organized as follows. \Cref{sec: methods} describes the basic expanding-box reduced magnetohydrodynamic (RMHD) 
model that we use throughout this work. We outline the useful ``wave-action'' form (\cref{subsub: wave action equations}), which facilitates analysis by factoring out the linear WKB wave evolution
brought about by expansion, before explaining the numerical method,  key parameters of the system, and the initial conditions used for the simulations. \cref{sec: basic evolution} then presents a brief overview of how the turbulence 
evolves, focusing on globally averaged quantities such as the energy, imbalance (normalized cross helicity), and residual energy. We will see that 
the evolution splits into two distinct phases, evolving from one  nonlinear solution of homogenous MHD (pure outwards propagating waves, high imbalance)  to 
another (magnetically dominated Alfv\'en vortices). 
In  \cref{sec: imbalanced phase} we examine the imbalanced phase, starting with a simple phenomenology based  on previous works \cite{Dmitruk02,verdini07,Chandran09a}  to understand the 
observed dynamics. We compare these phenomenological ideas
to the simulations' time evolution (\cref{sub: dmitruk decay}), spectra (\cref{sub: imbalanced spectra}), and frequency spectra (\cref{sub: anomalous coherence}), 
diagnosing how the suppression of wave collisions leads to ``anomalous coherence,'' enabling strong turbulence  despite the high imbalance. 
In \cref{sub: anastrophy}, we then examine the inverse transfer in detail, presenting a theoretical argument based on anastrophy  
to explain the observed results. 
The balanced, magnetically dominated phase is  examined in \cref{sec: Balanced phase}, starting 
with a focus on linear expansion-dominated (long-wavelength) physics (\cref{sub: linear balanced}). This linear physics controls the late-stage evolution 
of the system because the system self organizes to minimize its nonlinearity, explaining the strong dominance of magnetic over kinetic energy and various other features of its 
evolution (as well as a number of solar-wind observations). That 
this system does indeed morph into nonlinear solutions is  proved numerically (and argued theoretically) by directly fitting structures that grow in the
simulation (\cref{Subsec:Vortex}).

The paper contains a lot of detail about various aspects of the evolution. Therefore in \cref{sec: relation to observations}
we provide an extended summary of the observational relevance of our findings. 
This covers explanations of various existing observational results, such as the observed radial evolution and wind-speed dependence 
of imbalance and residual energy, as well as making  predictions that can 
be tested in future works to better understand the successes and limitations of the reflection-driven turbulence model. 
We conclude in \cref{sec: conclusion}.

\section{Methods}\label{sec: methods}
\subsection{The expanding reduced MHD model}\label{sub: expanding RMHD}
 
We wish to describe the turbulent dynamics of a plasma advected by an expanding wind and threaded by a mean magnetic field $\mB$ using the simplest possible formalism. We therefore assume that $\mB$ is radial, and that the fluctuations in the total field $\bm{B}$ and plasma velocity $\bm{u}$ are transverse and non-compressive, with characteristic scales well above the ion gyroscale (i.e., the fluctuations are polarized like shear-Alfv\'en waves).  We assume that the mean flow of the wind $\bm{U}$ is also radial, constant, and much larger than the Alfv\'en speed $\va\equiv|\mB|/\sqrt{4\pi\rho}$, where $\rho$ is the mass density of the plasma. These assumptions about $\boldsymbol{u}$ and $\bm{B}$
apply reasonably well to the solar-wind plasma in regions with $\mathcal{M}_{\rm A} \equiv |\bm{U}|/\va\gtrsim 1$ (i.e., beyond the Alfv\'en point) and where the Parker spiral is still well aligned with the radial direction  \cite{Parker1965}. Even with such simplifications, simulating such dynamics using an absolute frame of reference and over a large radial distance remains extremely costly in terms of computer power \cite{Perez13,vanballegooijen16,Chandran19}. We circumvent this difficulty by considering the turbulence dynamics in a frame co-moving with the spherically 
expanding flow---the so-called expanding box model (EBM) \cite{Grappin93}. 
Assuming that the domain is small compared to the heliocentric distance, the curvature of surfaces perpendicular to the radially expanding flow can be neglected, allowing the use of Cartesian coordinates and periodic boundary conditions in all three directions.  The resulting savings in numerical cost are redeployed to resolve the turbulence across a range of scales of unprecedented breadth.

These  approximations lead to equations that take the form of  standard ``reduced MHD'' (RMHD)  \cite{Kadomtsev73,Schekochihin2009}, with two modifications.
First, there appear additional linear terms proportional to $\bm{U}_{\perp}$, which is the part of the mean radial velocity perpendicular to the radial direction at the centerline of the simulation domain, which acts to expand the domain as it moves outwards  (note that the 
non-radial part of $\mB$ can be neglected because $|\bm{U}|\gg\va$ and due to the small spatial domain).  Second, the perpendicular gradient operator is modified to account for the increasing lateral stretching of the plasma
with distance: $\egrad \equiv (a^{-1}\partial _x,a^{-1}\partial_y, \partial_z)$,  where we use the local-box spatial coordinates $(x,y,z)$ and align the $z$ axis with 
the outwards radial direction at the centerline of the simulation domain.
Here $a$ is defined as the heliospheric distance $R$ of the co-moving frame, normalized by the initial radial distance $R_0$ (equivalently, it is  the perpendicular size of the domain):
\begin{equation}
    a(t)=\dfrac{R(t)}{R_{0}}=\dfrac{R_{0}+Ut}{R_{0}}=1+\dot{a}t,\label{Eq: a defn}
\end{equation}
where $\dot{a}=\partial a/\partial t = U/R_0$ is a 
constant for constant $\bm{U}$. Noting that  $\bm{U}_\perp = (\dot{a}/a)(x\xh+y\yh) $, one finds that the magnetic field, 
$\bm{B} =\mB+\bm{B}_{\perp}=B_z\zh+\bm{B}_{\perp}$, and the part of the perpendicular flow velocity that remains after the Galilean transformation, $\bm{u}_\perp = \bm{u} - \bm{U}$, evolve  as \cite{Grappin93}
\begin{align}
\dfrac{\rmd\bm{u}_{\perp}}{\rmd t} + \dfrac{\egradp p}{\rho}-\dfrac{\bm{B}\cdot\egrad\bm{B}_\perp}{4\pi \rho}=-&\bm{u}_{\perp}\cdot\egradp \bm{U}_{\perp}
 \nonumber \\=-&\dfrac{\dot{a}}{a}\bm{u}_\perp,
 \label{Eq:momentum}\\
 \label{Eq:induction}
\dfrac{\rmd\bm{B}}{\rmd t}-\bm{B}\cdot\egradp\bm{u}_{\perp}=
-\bm{B}\egradp & \cdot \bm{U}_{\perp} +\bm{B}_{\perp}\cdot\egradp \bm{U}_{\perp}
\nonumber \\ =- 2\dfrac{\dot{a}}{a} B_z \zh &-\dfrac{\dot{a}}{a} \bm{B}_\perp,
 \end{align}
where $\rmd/\rmd t = \partial/\partial t + \bm{u}_{\perp}\cdot\egradp$. The total pressure $p$, which includes both  magnetic and thermal pressures, cancels the compressive part of the nonlinear terms to enforce the incompressibility of the motions $\egradp \cdot \bm{u}_\perp=0$ \cite{Schekochihin2009}.  Defining 
the subscript $0$ to refer to a quantity at $t=0$ ($a=1$), conservation of mass and magnetic flux imply $
\rho=\rho_0/a^2$ and $B_z=B_{z0}/a^2$ (the latter being the solution of the $\zh$ component of \eqref{Eq:induction}), so that $\va =\vao/a$ \cite{Grappin93}. Note that because 
$\rho=\rho_0/a^2$, the perpendicular friction-like term in  \eqref{Eq:induction} associated with the spherical expansion ($-\dot{a}/a \bm{B}_\perp$) vanishes if one instead expresses the perpendicular magnetic field in velocity units $\bm{b}_{\perp}=\bm{B}_{\perp}/\sqrt{4\pi\rho}$ using $\partial_t \bm{b}_{\perp}=(4\pi\rho)^{-1/2}\partial_t \bm{B}_{\perp}+\dot{a}/a\, \bm{b}_{\perp}$. Because $\bm{u}_\perp$ is damped via $-\dot{a}/a\, \bm{u}_\perp$, this produces differential damping of the perpendicular magnetic and kinetic fluctuations during the radial transport.

The most important impact  of expansion  is that it causes Alfv\'enic  reflection. 
This can be seen by considering the Els\"asser variables $\bm{z}^{\pm}=\bm{u}_{\perp}\pm \bm{b}_{\perp}$,  which evolve as
\begin{equation}
 \dfrac{\partial \bm{z}^{\pm}_{\perp}}{\partial t}  \pm v_{A}\dfrac{\partial\bm{z}^{\pm}_{\perp}}{\partial z}+\bm{z}^{\mp}_{\perp} \cdot\egradp\bm{z}^{\pm}_{\perp}+\dfrac{\egradp p}{\rho}=-\dfrac{1}{2}\dfrac{\dot{a}}{a}\left( \bm{z}^{+}_{\perp}+\bm{z}^{-}_{\perp}\right).
 \label{Eq:Elsasser}
\end{equation}
We have taken $\mB$ to point in the negative radial direction ($B_z<0$ with $\va= |B_z|/\sqrt{4\pi\rho}$), so that $\bm{z}^{+}_{\perp}$ perturbations propagate outwards in the absence of reflection.
We see that the additional linear terms proportional to $\bm{U}_{\perp}$ appearing in Eqs.~\eqref{Eq:induction} and \eqref{Eq:momentum}  couple $\bm{z}^{+}_{\perp}$ and $\bm{z}^{-}_{\perp}$ perturbations through the final term in Eq.~\eqref{Eq:Elsasser},
with important consequences for their nonlinear evolution.

\subsubsection{``Wave-action'' form}\label{subsub: wave action equations}

It is convenient to rewrite  equations \eqref{Eq:Elsasser}  in terms of the so-called ``wave-action'' 
Els\"asser variables  \cite{heinemann80}, defined as 
\begin{equation}
\bzpm \doteq a^{1/2}\bm{z}^{\pm}_\perp \propto \frac{ \bm{z}^{\pm}_\perp }{\sqrt{\omegaa}}   , \label{Eq:wave_action} 
\end{equation}  
 where $\omegaa=k_{z}\va$ is the Alfv\'en frequency of a mode of wavenumber $k_{z}$. The second expression emphasizes the relationship to the  wave-action density \cite{Whitham1965}, which is $|\bm{z}^{\pm}|^{2}/\omegaa $ for a population of $\bm{z}^{\pm}$ fluctuations at some $k_{z}$, 
 and is conserved in the limit of high-frequency/short-wavelength waves. This highlights how the extra $a^{-1/2}$  factor compensates the decay of the $\bm{z}^{\pm}_{\perp}$ that arises because of the decreasing Alfv\'en frequency as the system expands, making $\bzpm$ the natural variables in which to consider turbulent-decay dynamics. Equations \eqref{Eq:Elsasser} then take the form,
\begin{equation}
 \dot{a}\dfrac{\partial \bzpm}{\partial a}\pm \va \dfrac{\partial \bzpm}{\partial 
 z}+\dfrac{1}{a^{1/2}}\left(\bzmp\cdot \egradp \bzpm +
 \dfrac{\egradp p}{\rho} \right)=-\dfrac{\dot{a}}{2a}\bzmp. \label{Eq:Equations}
\end{equation}
These equations can be equivalently derived from the ``flux-tube'' RMHD equations used by Refs.~\cite{Perez13,Chandran19} (see also \cite{verdini07,Chandran09a}) by identifying their $\bm{g}$ and $\bm{f}$ with $\bzp$ and $\bzm$, respectively, assuming $\va\ll U$ and $\eta\doteq\rho/\rho|_{U=\va}\ll1$, and
converting $\dot{a}\partial/\partial a $ in \eqref{Eq:Equations} into the derivative in the stationary frame $\partial /\partial t + U \partial /\partial R$.

For the remainder of this article we will usually use  wave-action variables with lengths and gradients defined in the co-moving frame, which does not change with $a$.
With this in mind, it is sometimes helpful to explicitly expand  $\egradp=a^{-1}\gradp$  and $\va=\vao/a$,
 in order to remove the hidden $a$-dependence of these terms in \cref{Eq:Equations}. Written in terms of $\ln a$, 
\cref{Eq:Equations} takes
 a form that is similar to standard RMHD in 
 a fixed-size domain, but with reflection terms and a time-variable coefficient $a^{-1/2} = e^{-\ln a/2}$ multiplying the nonlinear term: 
 \begin{equation}
\dot{a}\dfrac{\partial \bzpm}{\partial \ln a}\pm {\vao} \dfrac{\partial \bzpm}{\partial 
 z}+\frac{1}{a^{1/2}}\bzmp\cdot \gradp \bzpm + \gradp \tilde{p}=-\frac{\dot{a}}{2}\bzmp, \label{Eq:Equations lna}
\end{equation}
where $\tilde{p}=p/\rho$ enforces $\gradp\cdot \bzmp=0$.  It is often helpful to consider the turbulent evolution from the perspective of \cref{Eq:Equations lna}, 
multiplying lengths by $a$ and using \cref{Eq:wave_action} to convert back to physical quantities as need be. We similarly define wave-action velocities and magnetic fields, $\bu = a^{1/2} \bm{u}_{\perp}$ and $\bb = a^{1/2}\bm{b}_{\perp}$, respectively.

Throughout this article we use the tilde $\tilde{\cdot}$ to denote both wave-action-normalized fields and length scales defined in the co-moving frame (like $\gradp$). 
Because we have not transformed time in deriving \cref{Eq:Equations} or \eqref{Eq:Equations lna}, time-scales and frequencies are not denoted with a   tilde,
and can be equivalently defined in either the co-moving or physical frame with either wave-action or physical variables, as convenient. The same is  true for dimensionless quantities
and parallel length scales. 

\subsubsection{Conserved quantities}\label{subsub: conservation laws}

Unlike  homogeneous RMHD, individual wave-action Els\"asser energies $\tilde{E}^{\pm} \equiv \langle |\bzpm |^2 \rangle/4$ are not conserved in the presence of expansion. (Here and in the following,
angle brackets $\langle \dots \rangle$ denote a volume average over the expanding box in the co-moving frame.) The reflection terms can act as a source or a sink of wave-action energy, depending on the sign of the correlation between the Els\"asser fields, or residual energy $\tilde{E}^r =\langle \bzp\cdot\bzm\rangle/2 = \tilde{E}^{u} - \tilde{E}^{b}$ (we define also the wave-action kinetic and magnetic energies, $\tilde{E}^{u} = \langle |\bu|^{2}\rangle/2$ and $\tilde{E}^{b} = \langle |\bb|^{2}\rangle/2$, respectively). Specifically, one finds from \cref{Eq:Equations},
\begin{equation}
\dot{a}\dfrac{\partial \tilde{E}^{\pm}}{\partial a} =-\dfrac{\dot{a}}{4a}\langle \bzp\cdot\bzm\rangle=-\dfrac{\dot{a}}{2a}\tilde{E}^r\label{Eq: wave action energy def}
\end{equation}
In contrast, one sees that the reflection
sources  cancel out for the wave-action cross-helicity $\tilde{E}^{c}=
\tilde{E}^{+}-\tilde{E}^{-} = \langle \bu\cdot\bb\rangle$, which therefore remains, as in the homogeneous case, an ideal invariant \cite{Verdini08}, 
\begin{equation}
\dfrac{\partial \tilde{E}^{c}}{\partial a} =0.\label{eq: wave action cross helicity}
\end{equation}
We note that although the fluctuation energy is not conserved, one can show using the full system of equations (without making the expanding-box approximation) that  the energy gained or lost by the fluctuations is compensated by an equal and opposite change in the energy of the background flow, 
with \cref{Eq: wave action energy def,eq: wave action cross helicity} resulting from total energy and cross-helicity  conservation 
under appropriate assumptions \cite{Chandran15,Perez2021}.
It is also helpful to define the total energy $\tilde{E} = \tilde{E}^{+}+\tilde{E}^{-}=\tilde{E}^{u}+\tilde{E}^{b}$.

 \subsection{Numerical method and setup}
 
 Taking advantage of the periodic boundary conditions, we solve \cref{Eq:induction,Eq:momentum} (or equivalently, \cref{Eq:Elsasser}, \eqref{Eq:Equations}, or \eqref{Eq:Equations lna}) with a modified version of the Fourier pseudo-spectral code TURBO \cite{Teaca09}.
 We advance in time with a third-order modified Williamson algorithm (a four-step, low-storage Runge–Kutta method \cite{Williamson80})  for the nonlinear terms and implicitly evaluate the linear terms exactly. The simulation domain is a cube of size $L_\perp=L_z=2\pi$  with a resolution 
$n_{\perp}^{2}\times n_{z}$. Note that the system \eqref{Eq:Equations} has a rescaling symmetry, whereby all relative fluctuation amplitudes can be arbitrarily rescaled as long as the ratios of all perpendicular to parallel scales are rescaled by the 
same amount. Therefore, the parallel and perpendicular units of length are independent. The code units are set by this and by $v_{A0} = 2\pi$. 
Nonlinear terms are partially dealiased using a phase-shift method \cite{Patterson71}. The main simulations presented below will use a spatial  resolution of $n_{\perp}^{2}\times n_{z}=1536^{2}\times 128$ for the full simulation evolution, but are refined to  $n_{\perp}^{2}\times n_{z}=8192^{2}\times 256$ 
around specified radii of interest and allowed to evolve briefly, in order to resolve spectra at smaller scales.

We add a form of dissipation (``hyperviscosity'')  
\begin{equation}
-\nu^{\pm}_{\perp}\egradp^{6}\bzpm -\nu^{\pm}_{z}\partial_{z}^{6}\bzpm
\label{Eq:Hyper-visc}
\end{equation}
 to the right-hand side of \cref{Eq:Equations}  to absorb the turbulent energy at small scales. The hyper-viscosity coefficients $\nu^{\pm}_{\perp}$ and $
\nu^{\pm}_{z}$ are adaptive, \textit{viz.,} they are re-evaluated at each time step to ensure that dissipation occurs near the smallest scales of the grid in order to maximize the inertial range. This is 
is necessary because the turbulent amplitudes change by orders of magnitude over the course of the simulations, thus changing the dissipation 
scale for a given (fixed) hyperviscosity significantly. The method is explained in more detail \cref{app: adaptive viscosity}.

 \subsubsection{Simulation parameters}
 
 In the expanding RMHD equations, there are three  ratios of timescales that will prove important 
 for the dynamics. We will define these in more detail below, but feel it useful to introduce the notation here:  $\chi_{\rm A}$ will denote the usual ratio of Alfv\'enic to nonlinear timescales \cite{Goldreich95,Mallet2015}; $\chi_{\rm exp}$, the 
 ratio of the expansion to nonlinear timescales; and $\Delta = \chi_{\rm exp}/\chi_{\rm A}$, the ratio 
 of the expansion to Alfv\'enic timescales. 
 Because of the rescaling symmetry of the RMHD equations, aside from resolution and dissipation properties, 
 two of these three parameters set the important parameters of a given simulation. 
It is most natural to set $\chi_{\rm exp }$ and $\chi_{\rm A}$ via the initial conditions 
 (discussed below) and fix the ratio of the box-scale Alfv\'en frequency ($\omega_{\rm A,box}=  2\pi\va/L_{z}$) to the expansion rate,
 \begin{equation}
\Delta_{\rm box} \doteq \frac{\omega_{\rm A,box}}{\dot{a}/a}=\dfrac{2\pi}{L_z}\dfrac{\vao}{\dot{a}}.   \label{Eq: Delta in sim}
\end{equation}
Note  that $\Delta_{\rm box}$ remains constant throughout the evolution because $\va\propto 1/a$. In all simulations, we will take $\Delta_{\rm box}=10$.
Using $\dot{a}/a = U/R$ (see Eq.~\eqref{Eq: a defn}), this implies that the box has physical size
\begin{equation}
    L_z = \frac{2\pi}{10} R \frac{\vao}{U} \approx 5.1\times 10^6{\rm km}\, \frac{\mathcal{M}_{\rm A}}{3} \frac{R}{35 R_\odot},  
\end{equation}
where we have chosen physical values that are characteristic of early Parker Solar Probe passes \cite{Bale2019}.
This scale corresponds to structures that are advected past the spacecraft at frequency $f=U/L_z \approx 5.9\times 10^{-5}{\rm Hz}\, (R/35 R_\odot)^{-1} (\mathcal{M}_{\rm A}/3)^{-1}$, which is below the observed correlation scale of the turbulence, as desired. Because $\mathcal{M}_{\rm A}\propto 1/R$ and $U$ is constant 
in the super-Alfv\'enic wind, this minimum resolved frequency remains constant as the simulation evolves, although the  correlation scale is observed to increase.

Note that the choice of $\Delta_{\rm box}$ can be equivalently understood as setting the resolution in $k_z$ of the simulation:  with infinite spatial resolution, a longer box, which contains 
lower $k_{z}$ modes, is identical to a shorter box with smaller $\Delta_{\rm box}$. It is also of note that  
there exist  $k_{z}=0 $
two-dimensional modes, which do not propagate, unlike the other modes in the box. While these are, 
in some respects,  an artefact of the expanding box's periodic boundary conditions, we argue below that 
they are capturing important physical effects and should not be artificially excluded (see \cref{sec: Balanced phase}).  

\subsubsection{Initial conditions}

Rather than realistically simulate a patch of  solar wind as it propagates outwards, the goal of this work is to distill and understand theoretically the key physical features
of reflection-driven turbulence. Therefore, our initial conditions are idealized and designed to understand the model itself, on the assumption that this is a prerequisite for understanding the physical processes it attempts to represent. Anticipating the result that the correlation scales of the turbulence will increase significantly as it evolves, 
it is thus important to start with fluctuations on scales well below the box scale in order to avoid artificially constraining the system's evolution.  
We choose to obtain the initial $\bzp$ fields from a balanced  RMHD simulation evolved into its statistically stationary turbulent state, 
loosely motivated by the idea that outwards Alfv\'enic fluctuations could ``escape'' into the corona through an effective high-pass filter from a region of nearly balanced stronger turbulence \cite{vanBallegooijen2011} (although, of course, the EBM is  formally valid only outside the Alfv\'en point by which point the turbulence will have evolved \cite{Perez13,vanballegooijen16}). 
 The forcing of this balanced simulation  is local in Fourier space, acting on all the modes within the ring $k_{\perp} \in 2\pi/L_{\perp} \left[ 99.5, 100.5 \right]$ and $\vert k_z \vert=  2\pi/L_{z}$, and is designed so as to keep the rate of injection of energy constant with the amplitude needed for critical balance \cite{Goldreich95}.  
 This creates  initial fluctuations with a correlation scale modestly above the forcing scale, with a perpendicular correlation length $L_+\approx L_{\perp}/75$ and parallel correlation length ${\approx}L_{z}$. In the infra-red range (scales larger than the perpendicular correlation length), the initial energy spectrum scales approximately as ${\propto} k_\perp$ in accordance with  theoretical expectations \cite{Schekochihin2022}.  We use the $\bzp$ field thus obtained to initialize the $\bzm$ one by setting $\bzm=-\kappa\bzp$ with $\kappa$ such that $1-\sigma_c=\num{1e-4}$ (this choice is not of great importance because the system rapidly self adjusts).
 
 Given this choice of a spectrum of fluctuations, the  only remaining parameter of interest is the RMHD fluctuation amplitude, which, as shown 
 below, has a strong impact on the turbulence evolution. Given the
 rescaling symmetry discussed above, this amplitude should be thought of as controlling the ratio of the nonlinear timescale $\tau_{\rm nl}^{\mp}\sim (k_{\perp} z^{\pm})^{-1}=a^{-3/2}(\tilde{k}_{\perp}\zp)^{-1}$
 to the linear timescales $(k_{\|}\va)^{-1}$ and $a/\dot{a}$, as opposed to directly 
 setting the physical turbulent amplitude $z^{+}/\va$ (or $|\bm{B}_{\perp}|/\overline{B}$ or $ |\bm{u}_{\perp}|/\va$).
 Accordingly, we set 
 \begin{equation}
\chi_{\rm exp0} \doteq \frac{k_{\perp0} z^{+}_{\rm rms0} }{\dot{a}/a} \:\text{ and } \:\chi_{\rm A0} \doteq \frac{k_{\perp0} z^{+}_{\rm rms0} }{k_{z0} \va}
\end{equation}
as simulation parameters by rescaling $\bzp$ by the required amount. Here
 $k_{\perp0}$ and $k_{z0}$ are the initial inverse correlation lengths, $z^{+}_{\rm rms0}$ is the initial root-mean-square (rms) fluctuation amplitude, 
 and the ratio $\chi_{\rm exp0}/\chi_{\rm A0}=\Delta_{\rm box}$ is fixed to be 10 for all simulations as described above (i.e., rescaling $\bzp$ sets both $\chi_{\rm A0}$ and $\chi_{\rm exp0}$
 because we have already fixed $\Delta_{\rm box}$). We shall see that because they have stronger nonlinearity, simulations with larger $\chi_{\rm exp 0}$ remain 
 in the strongly nonlinear regime for longer,  thus displaying more clearly the relevant power-law behavior and clarifying the
 analysis. Most figures and discussion will thus focus on the highest-$\chi$ case run, which has $\chi_{\rm exp0}=960$ ($\chi_{\rm A0}=96$) and a resolution $n_\perp^2\times n_z=1536^2\times 256$. This value of $\chi_{\rm exp0}$ 
 is rather large compared to the solar wind around $R_{\rm A}$ at the correlation scale of the turbulence (see \cref{sec: relation to observations}) but useful nonetheless for understanding the key physics. 
 We have run a series of simulations down to $\chi_{\rm exp0}=0.75$ and will show some of these for comparison.

 This method of constructing initial conditions, while straightforward and well controlled,  has the downside of placing the plasma into an artificial ``super-critically balanced'' state ($\chi_{\rm A}\sim k_{\perp}z^{+}/k_{\|}\va>1$). The consequence is that,
 over a relatively short transient initial phase as the fields start evolving nonlinearly, neighboring  planes along the $\hat{\bm{z}}$ direction decorrelate and 
  develop small parallel scale fluctuations until $\chi_{\rm A}\sim 1$, establishing critical balance. This transient process generates a flat $k_\|$ spectrum (white noise) up to the parallel scale at which $k_\| \va$ balances the nonlinear mixing, which is the Fourier-space hallmark of critical balance \cite{Schekochihin2022}. 
  This process occurs over a timescale comparable to the nonlinear time at each scale, which is rapid compared to the time it takes the system to decay and change regimes, 
  so we believe this choice does not strongly impact our results. However, future work should explore the effect of this choice, other initial conditions, and  $\Delta_{\rm box}$  in more detail in order  better understand the impact of our choices.

\begin{figure*}
\centering
\includegraphics[width=1\linewidth]{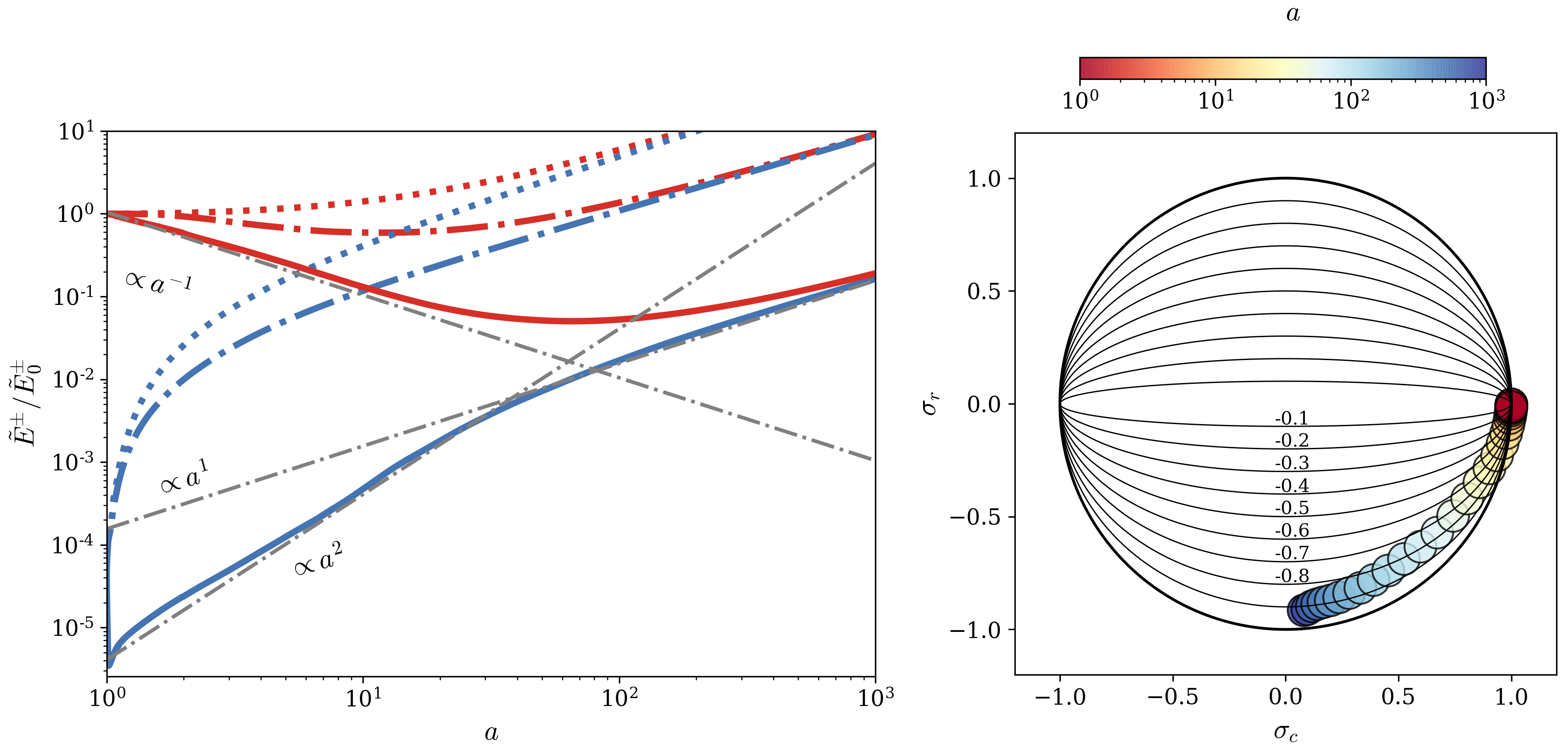}
\caption{ Left panel: Radial evolution of wave action energies $\tilde{E}^+$ (red lines) and 
$\tilde{E}^-$ (blue lines) for  three  simulations with different amplitude initial conditions. 
Solid 
lines show our highest-amplitude  $\chi_{\rm exp0}=960$ ($\chi_{\rm A0}=96$) simulation, dash-dotted lines show the $\chi_{\rm exp0}=7.5$ ($\chi_{\rm A0}=0.75$) simulation, and dotted 
lines show the $\chi_{\rm exp0}=0.75$ ($\chi_{\rm A0}=0.075$) simulation in the weak regime. We normalize each curve to its initial $\tilde{E}^+$ to facilitate comparison and the dotted-grey lines indicate various power laws for reference (see text).
 Right panel: Parametric representation of $\sigma_r$ and $\sigma_c$ during the evolution of the $\chi_{\rm A0}=96$ simulation. 
 The colors (on a logarithmic scale) indicate the 
normalized radial distance $a$. Solid lines represent contours of constant $\sigma_\theta$  as labelled (see text). \label{Fig: time evolutions}}
\end{figure*}

\begin{figure*}
 \centering
 \includegraphics[width=0.3\linewidth]{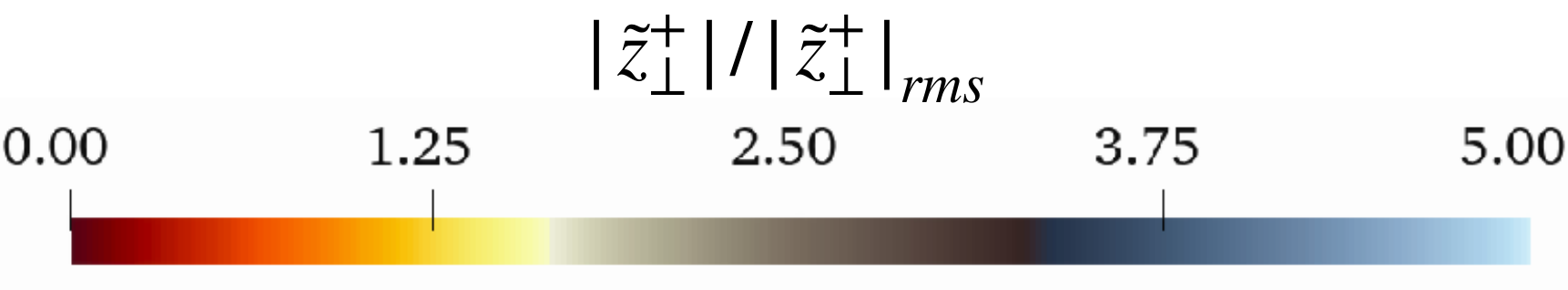}
  \hspace{1.25cm}
  \includegraphics[width=0.3\linewidth]{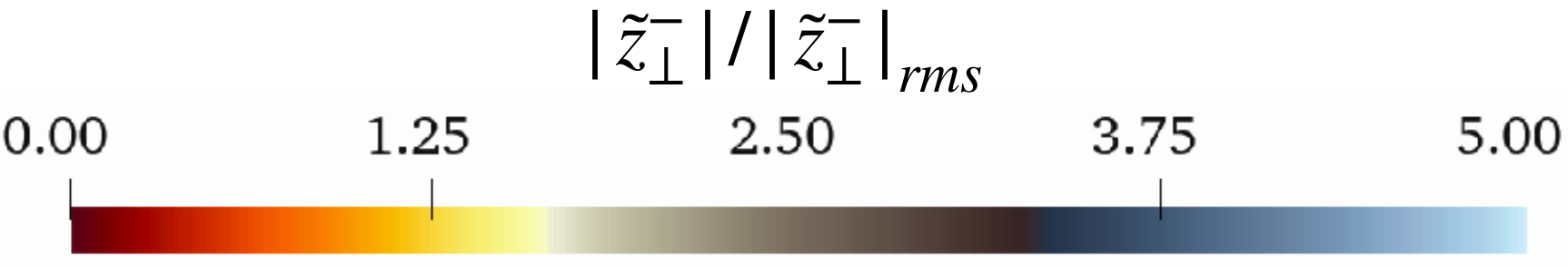}
   \includegraphics[width=0.77\linewidth]{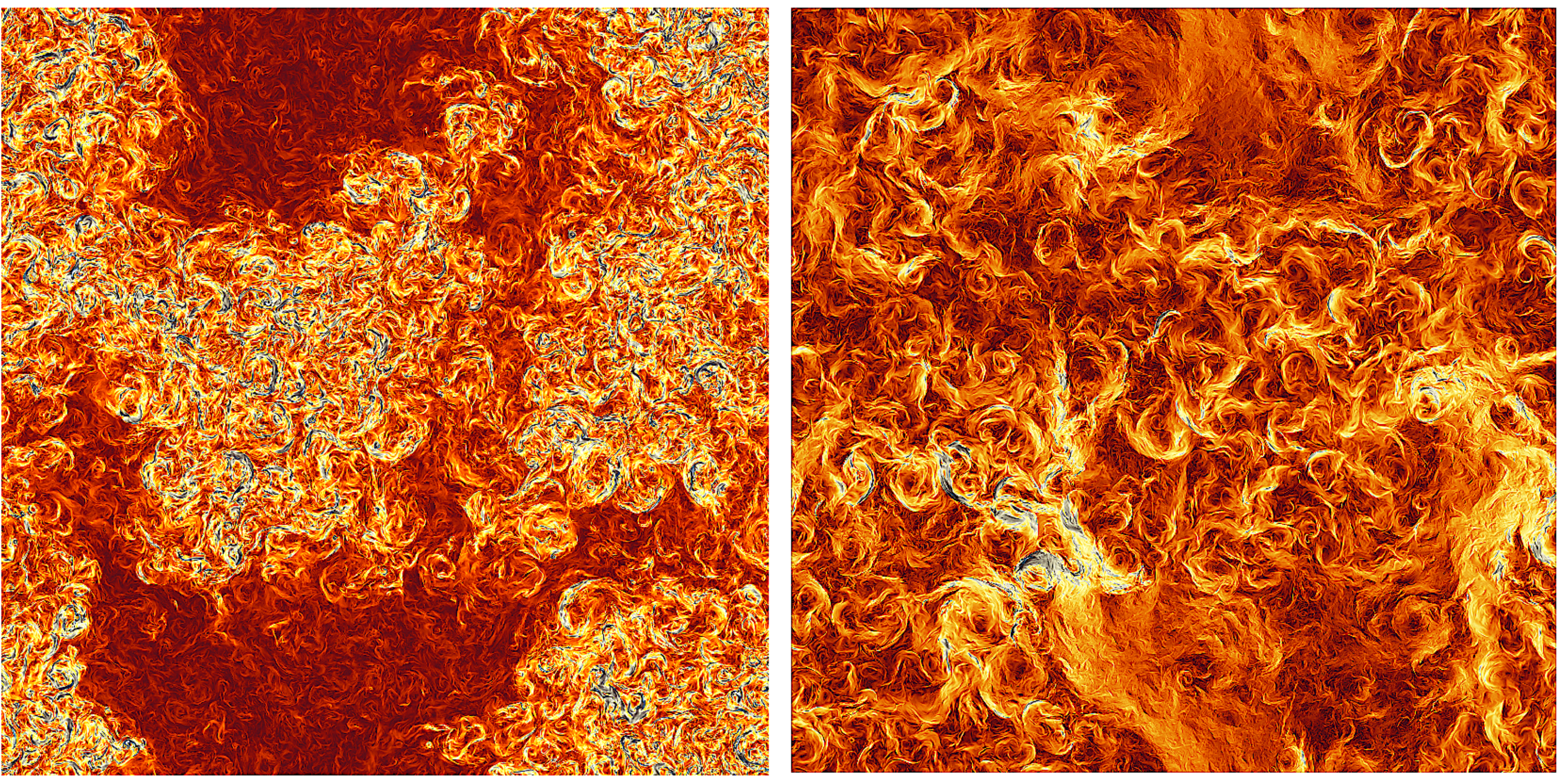}
   \includegraphics[width=0.77\linewidth]{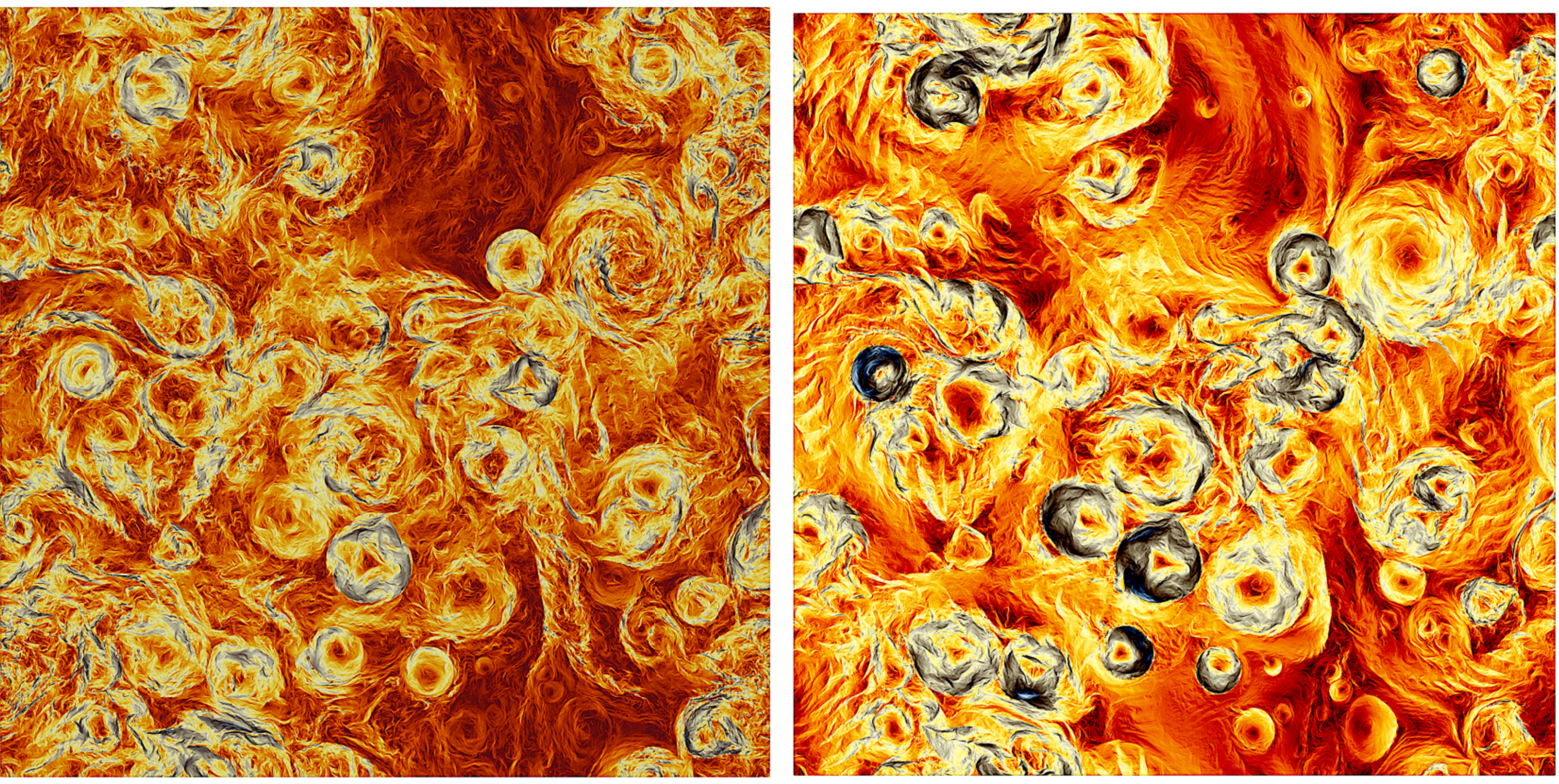}
   \includegraphics[width=0.77\linewidth]{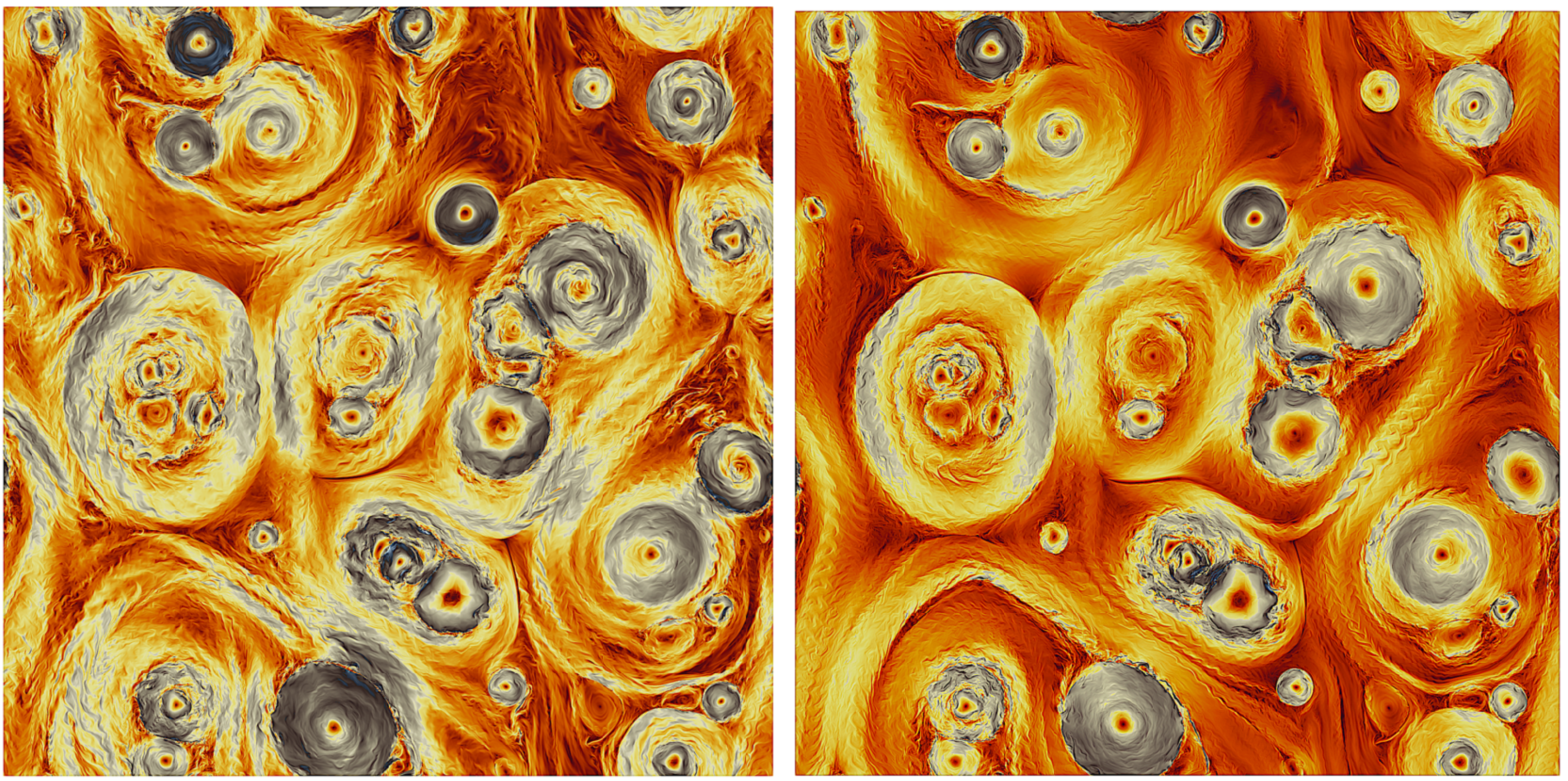}
   \caption{Snapshots of the Els\"asser fields $|\bzp|$ (left panels) and $|\bzm|$ (right panels) in the plane perpendicular to the mean magnetic field for three different  radial distances.   The top panels illustrate $a=5$ during the imbalanced  decay phase; the   middle panels show $a=50$, which is shortly before the transition to the balanced phase; the bottom panels show $a=250$ in the balanced, magnetically dominated regime. This simulation has a resolution of $n_\perp^2\times
   n_z=8192^2\times 256$ and is initialised by progressively refining the $n_\perp^2 \times n_z= 1536^2 \times 256$ simulation that was run 
   from $a=1$ to $a=1000$.}
    \label{Fig:Snapshot}
\end{figure*}

 \section{Basic evolution}\label{sec: basic evolution}

Starting from the initial conditions described above, we evolve the system with $a$ (equivalently, with time), up to $a=1000$. While 
this would correspond, in principle to an extremely large physical radius [$R=1000R_{\rm A}\approx 70 {\rm AU} (R_{\rm A}/15 R_{\odot})$], 
we reiterate that we are deliberately exploring more extreme parameters in order to better 
characterize the physics of reflection-driven turbulence. For more realistic initial conditions with lower $\chi_{\rm exp0}$, the behavior and transitions we describe below will occur at smaller $a$.

As illustrated in  \cref{Fig: time evolutions}, which shows important aspects of how simulations with different initial amplitudes ($\chi_{\rm A0}$) evolve with $a$,  the system's evolution is naturally divided into two distinct phases, discussed separately in \cref{sec: imbalanced phase} and \cref{sec: Balanced phase} below. 
Following a short initial transient, when $\bm{z}^{-}$ and the parallel scales rapidly adjust (see above), the  first ``imbalanced'' phase 
involves turbulence where the nomalized cross helicity, or imbalance,
\begin{equation}
 \sigma_c \doteq \dfrac{\tilde{E}^{+}-\tilde{E}^{-}}{\tilde{E}^{+}+\tilde{E}^{-}} = \frac{\tilde{E}^{c}}{\tilde{E}},
 \end{equation}
is almost maximal (unity), as in the initial conditions. In the strong nonlinear regime ($\chi_{\rm A0}=96$; solid lines in the left panel of \cref{Fig: time evolutions}), the  turbulent energy decays as $\tilde{E}\approx \tilde{E}^{+}\propto a^{-1}$, signalling turbulent heating of the plasma.  In contrast, in the second ``magnetically dominated'' or balanced phase, which starts at around $a\approx80$ for the $\chi_{\rm A0}=96$ simulation, 
$\sigma_{c}$ approaches $0$ with $\tilde{E}^{+}\approx \tilde{E}^{-}$, and surprisingly, $\tilde{E}$ starts growing in time. 
This is a consequence of the system developing  a large negative
normalized residual energy,
\begin{equation}
\sigma_r\doteq\dfrac{\tilde{E}^{u}-\tilde{E}^{b}}{\tilde{E}^{u}+\tilde{E}^{b}} = \frac{\tilde{E}^{r}}{\tilde{E}},
\end{equation}
which, as seen from \cref{Eq: wave action energy def}, can cause  $\tilde{E}$ to grow as $\tilde{E}\propto a$ (as observed) in the absence of dissipation. 
We show this evolution graphically with the ``circle plot'' in the right-hand panel of \cref{Fig: time evolutions}. This illustrates 
 the  evolution of $\sigma_c$ and $\sigma_r$  during the radial transport  \cite{Bruno07}, which are constrained by the relationship between $\tilde{E}^\pm$, $\tilde{E}^u$, and $\tilde{E}^b$  to lie within the circle  $\sigma_c^2+\sigma_r^2=1$.
The fact that the evolution remains near the edge of the circle 
indicates that the fields maintain a high level of ``Els\"asser alignment'' between $\bzp$ and $\bzm$, with 
\begin{equation}
\sigma_{\theta}\doteq \dfrac{\langle \bzp\cdot\bzm \rangle}{\langle |\bzp|^{2}\rangle^{1/2}\langle|\bzm|^{2}\rangle^{1/2}}=\dfrac{\sigma_{r}}{\sqrt{1-\sigma_c^{2}}},
\label{Eq:Cos}
\end{equation} 
close to  $-1$ in the later stages of the simulation (the isocontours of $\sigma_{\theta}$ are shown by solid lines in \cref{Fig: time evolutions}).
This strong alignment is likely primarily a consequence of the reflections, which generate $\bzm$ fluctuations that are perfectly aligned with $-\bzp$, although the mutual shearing of the Els\"asser fields is also known to generate aligned fluctuations even  in 
homogeneous Alfv\'enic turbulence \cite{Chandran15}.
The simulation's evolution bears  a striking resemblance to the joint distribution of normalized cross helicity and residual energy observed in highly Alfv\'enic fast-solar-wind streams \cite{Bruno07,Wicks13,DAmicis2021}, 
providing good evidence that, despite the drastic approximations involved with our model, it captures some of the key physics of solar-wind turbulence.

The properties of the turbulence change dramatically between the two phases, as illustrated by the perpendicular  snapshot of $\bzpm$ shown in  \cref{Fig:Snapshot}.
Most obviously, the turbulence dramatically increases in scale with time, starting from the very small scales of the initial conditions (top panels) to reach nearly 
the box scales by the latest times (bottom panels). We will argue below that 
this is a consequence of the anomalous turbulent growth of ``wave-action anastrophy'' during the imbalanced phase, which significantly 
constrains the turbulence as it decays, forcing it to rush to larger scales and form a split cascade. At early times, the structures in $\bzp$ and $\bzm$ are rather different, with different dominant scales, but 
as the turbulence enters the magnetically dominated phase (middle panel) the two become more similar as it becomes balanced. A key
change (not shown in  \cref{Fig:Snapshot}), is that the turbulence becomes more two-dimensional at larger $a$, with structures 
across a wide range of $k_{z}$ at earlier times (top panel) giving way to predominantly $k_{z}=0$ modes by the $a=250$ snapshot shown in the bottom panel. 
While true $k_{z}=0$ modes are of course an artefact of the periodicity of the EBM, their key feature as pertains to reflection turbulence is that
they are expansion dominated and do not propagate, unlike Alfv\'en waves. Since this is the case for any sufficiently long-wavelength mode, even in non-periodic settings or the real solar wind (specifically, those with  $\Delta = k_{z}\va/(\dot{a}/a)<1/2$; see \S\cref{sec: Balanced phase}), we 
argue that these dynamics are physical and likely have already been observed in the solar wind. 
As seen also in the left panel of \cref{Fig: time evolutions}, there is little turbulent heating in this phase, which (we will show) occurs because the circular structures  approach local nonlinear ``Alfv\'en vortex'' solutions \cite{Petviashvili92,Alexandrova08}, which slows down  their evolution significantly, impeding their dissipation.

We now explore the two phases in more detail, attempting to diagnose and understand key features of their turbulence to make detailed predictions
for solar-wind observations.

\section{Imbalanced phase}\label{sec: imbalanced phase}

In this section, we explore the turbulence in the imbalanced phase of the simulations, which applies when $\zp\gg \zm$, for $a\lesssim 50 $ in the $\chi_{\rm A0}=96$ simulation   (see \cref{Fig: time evolutions}). Based on \cref{Fig: time evolutions,Fig:Snapshot}, the key features of this phase that we wish to understand are (i) the power-law evolution of 
$\tilde{E}^{\pm}$, which sets the overall heating (turbulent-decay) rate as a function of radial distance, and (ii) the cause of the significant increase in the fluctuations' scale during their evolution. 
To interpret the basic time evolution of $\tilde{E}^{+} $ and $\tilde{E}^{-}$, we first (\cref{sub: dmitruk decay}) review and assess  phenomenological ideas based on Ref.~\cite{Dmitruk02},  which have been used in a number of previous 
works to predict and understand reflection-driven turbulence both in- and outside the Alfv\'en point \cite{verdini07,verdini10,Chandran09a,Chandran19}.
While the phenomenology is consistent with some general features of the observed time evolution (\cref{sub: dmitruk decay}) and spectra (\cref{sub: imbalanced spectra}),  we will find some important differences that we cannot, at this point, satisfactorily 
explain.  Whether these signal fundamental issues with the theoretical basis of the model, or just more minor discrepancies, remains unclear.
In this discussion, we will see that feature (ii) (the rapid increase in the the scale of the fluctuations) happens to not influence the decay, so it can be discussed separately. We argue in \cref{sub: anastrophy}
that this feature arises from the surprising property of anomalous turbulent ``wave-action anastrophy'' growth, which constrains the dynamics and forces $\zp$ to rush to large scales as it decays 
via a split cascade.

\subsection{Turbulent decay phenomenology}\label{sub: dmitruk decay}

The basic idea of the phenomenological model is to treat the dominant $\zp$ fluctuations as a standard decaying-turbulence problem, while $\zm$ is effectively strongly forced by reflection and damped by turbulence. In more detail, 
because $\tilde{E}^{+} \gg \tilde{E}^{r}$ when $\tilde{E}^{+}\gg \tilde{E}^{-}$ (as assumed), reflection is negligible for the $\zp$ field, and consequently, for the forcing/damping of the wave-action 
energy (see Eq.~\eqref{Eq: wave action energy def}). This implies $\tilde{E}^{+}$ is approximately ideally conserved during this phase and its turbulent decay occurs only due to 
 non-linear coupling with $\zm$. 
 Throughout this phase, the $\zm$ fluctuations, which are forced by reflections, remain very low amplitude; therefore \textit{a-priori}, one might expect $\zp$ fluctuations to be in the weak regime. 
 However, we assume \cite{Velli89,Lithwick07,Chandran19,Schekochihin2022},  providing detailed numerical justification below (\cref{sub: anomalous coherence}), 
 that the $\zm$ fluctuations remain ``anomalously coherent'' with the $\zp$, because their forcing via reflection is highly coherent (${\propto}-\bzp$) thus ``dragging''
 $\zm$ along with the $\zp$ in time.
 The consequence is twofold: first, by moving  into the frame that propagates outwards with  $\bzp$,
it allows one to ignore the Alfv\'enic propagation terms for both $\bzp$ and $\bzm$; second, it allows the estimation of turbulent cascade times using the standard nonlinear turnover times (unlike for weak turbulence).
Therefore, the turbulent decay time $\tau^{\pm}$ of $\zpm$ is
\begin{equation}
\tau_{\mp}^{-1}\sim a^{-3/2}\frac{\zpm}{\lampm} = \frac{z^{\pm}}{\lambda^{\pm}},
\end{equation}
where $\lampm$ are the characteristic perpendicular scales of $\zpm$ in the co-moving frame that govern the decay/growth of $\zmp$, and $\zpm$ represents the rms amplitude of $\bzpm$. Variables 
without the tilde are in the physical frame with physical units,  showing how the $a^{-3/2}$ factor arises from the use of wave-action variables.

Based on these assumptions, we compute the evolution of   $\zm$  via the balance of reflection and nonlinear decay, ignoring the Alfv\'enic and  time-evolution terms (the latter is small, as justified below). The evolution of $\zp$ results  from its nonlinear turbulent decay via the $\zm$ that it has sourced.
The scheme then yields the following phenomenological evolution equations for $\tilde{E}^{\pm}$ \cite{Dmitruk02}:
\begin{subequations}
\begin{gather}
\dot{a}\dfrac{\partial  {\tilde{E}^{+}}}{\partial a} \sim  
-\dfrac{1}{a^{3/2}}\dfrac{\zm}{\lamm} \tilde{E}^{+},\label{Eq: heuristic energy +} \\
\dfrac{1}{a^{3/2}}\dfrac{\zp}{\lamp}\tilde{E}^{-}\sim \dfrac{\dot{a}}{a}|\tilde{E}^{r}|\sim \frac{\dot{a}}{a}|\sigma_{\theta}| \zp \zm.
\label{Eq: heuristic energy -}\end{gather}\label{Eq: heuristic energy}\end{subequations}
Writing \eqref{Eq: heuristic energy -} for $\zm$ instead gives 
\begin{equation}
\zm \sim \dot{a}a^{1/2} \lamp|\sigma_{\theta}|, \label{Eq:Minus_Law}
\end{equation}
whereby we see the interesting feature that the  amplitude of $\zm$ is independent of that of $\zp$ (other than indirectly through $\lamp$ and $\sigma_{\theta}$).
This occurs because $\zp$ acts to both drive and dissipate the $\zm$ energy. This independence from the $\zp$ spectrum also suggests that, with various caveats discussed below (\cref{sub: imbalanced spectra}), 
it could be reasonable to reinterpret the balance of reflection and nonlinear damping as applying at each scale separately, thus replacing the  $\lamp$ in \cref{Eq:Minus_Law} with $\tilde{k}_{\perp}^{-1}$ and
making $\zm$ the rms amplitude of the $\zm$ increment across a distance $\tilde{k}_\perp^{-1}$ in the perpendicular plane. This gives $\zm(k_{\perp})\propto k_{\perp}^{-1}$, or a ${\propto} k_{\perp}^{-3}$ energy spectrum for $\zm$.
We can insert \cref{Eq:Minus_Law} into \cref{Eq: heuristic energy +} to obtain the total energy ($\tilde{E}\approx \tilde{E}^{+}$) decay,
\begin{equation}
\frac{\partial \ln \tilde{E}^{+}}{\partial a} \sim -\frac{1}{a} \frac{\lamp}{\lamm}\sigma_{\theta}.\label{eq: plus decay law}
\end{equation}

Several other points are worth noting. First, 
the anomalous coherence will break down once the effect of  $\zp$  on $\zm$ enters the  weak regime
(in which case $\zm$ can propagate away from its $\zp$ source). The phenomenology thus requires  
\begin{equation}
\chi_{\rm A}\doteq \frac{(\taum)^{-1}}{ \va/\ell_{\|}} \sim \frac{\zp /\lamp}{a^{1/2}\vao/\ell_{\|}}\gtrsim1\label{eq: chi exp def}
\end{equation}
where $\ell_{\|}$ is the parallel correlation length ($\chi_{\rm A}>1$ may be unphysical for other reasons, but the phenomenology itself is fundamentally 2D, ignoring $\ell_{\|}$).
Second,  we verify that the neglect of $\partial_{t} \zm $ is consistent, so long as anomalous coherence allows us to ignore the Alfv\'enic propagation of $\zm$ in the frame of $\zp$, by noting that $\partial_{t} \zm  \sim (\dot{a}/a)\zm$ is 
a factor ${\sim}\zm/\zp$  smaller than the reflection term in \cref{Eq: heuristic energy -}. 
Third, there exists an additional constraint implicit in \eqref{Eq: heuristic energy}, which comes from noting that \cref{Eq:Minus_Law} is equivalent to 
\begin{equation}
\zm \sim \frac{\zp}{\chi_{\rm exp}},\label{eq: first chi exp}\end{equation}
where \begin{equation}
\chi_{\rm exp}\doteq \frac{(\taum)^{-1}}{\dot{a}/a} \sim \frac{\zp /\lamp}{a^{1/2}\dot{a}}\label{eq: chi exp def}
\end{equation}
 is the ratio of the nonlinear damping to reflection rates. Thus, 
the phenomenology can only be valid for sufficiently large-amplitude $\zp$ with  $\chi_{\rm exp}\gg1$, irrespective of the fluctuation's parallel scale, and we expect the transition to the balanced regime to occur when $\tilde{E}^{+}$ decays sufficiently so that $\chi_{\rm exp}\sim 1$.
$\chi_{\rm exp}$ will feature prominently below as the key parameter that controls the transition from the imbalanced to balanced phase.

Previous treatments \cite{verdini07,Chandran09a,verdini10} have taken $\lamp$ and $\lamm$  in Eqs.~\eqref{Eq: heuristic energy} to
be the same and constant in time in  the co-moving frame. But, the argument about the $\zm$ balance and spectrum in  the previous paragraph, as well as  
decaying turbulence theory in general \cite{Kolmogorov41}, suggest that there is no reason to expect this to be the case. Indeed, if the $\zm $ spectrum
was ${\propto}k_{\perp}^{-3}$ as suggested above, then --- irrespective of the dominant scales of $\zp$ --- the correlation scale of $\zm$ 
would become the largest scale at which the arguments leading to 
Eqs.~\eqref{Eq: heuristic energy} break down (e.g., where $\chi_{\rm exp}<1$, or where the turbulence becomes weak). In addition, 
we will show below that the co-moving scales of $\zp$ evolve in time as a result of another nonlinear conservation law  (that for the ``wave-action anastrophy''). 
Herein lies the problem that complicates the comparison of the phenomenology to the numerical experiments: it is not clear
what constrains  the $\lamp$ and $\lamm$ scales in Eqs.~\eqref{Eq: heuristic energy}, but their time evolution is crucial 
for determining many key aspects of the turbulent evolution. In addition, it is not clear how the evolution of $\lamp$, which is the characteristic scale of $\zp$ that controls the nonlinear evolution of $\zm$,  relates to that of the correlation scale $\Lp$ of $\zp$. This  allows us to consider the evolution of $\Lp$ separately from the decay phenomenology, unlike in standard
decaying turbulence theory (\cref{sub: anastrophy}), but the cause of this apparent discrepancy between $\lamp$ and $\Lp$ remains a poorly understood aspect of the phenomenology.

\subsubsection{Numerical results}

Consider first the left-hand panel of  \cref{Fig: time evolutions},
focusing on the decay (growth) of $\tilde{E}^{+}$ ($\tilde{E}^{-}$) for the $\chi_{\rm A0}=96$ ($\chi_{\rm exp0}=960$) simulation (solid lines), which undergoes a long period 
of power-law evolution before reaching the balanced regime. 
We see that $\tilde{E}^{-}\propto a^{2}$, which is significantly faster than the simplest  prediction from \cref{Eq:Minus_Law} with $\lamp\sigma_{\theta}\sim{\rm const.}$  (yielding $\tilde{E}^{-}\propto a^{1}$). While this is perhaps not surprising, since, as seen in \cref{Fig:Snapshot}, the fluctuations' scales are increasing rapidly with 
time (thus presumably increasing $\lamp$), we have not identified a clear candidate for providing the additional factor of $a^{1/2}$ in \cref{Eq:Minus_Law}.\footnote{Intriguingly, the correlation 
length of the residual energy, which is the forcing scale of $\zm$ and could perhaps be heuristically identified with $\lamp\sigma_{\theta}$, grows  as approximately ${\propto}a^{1/2}$, providing
 a good match to the observed growth of the amplitude of $\zm$ from \cref{Eq:Minus_Law} in some simulations. However, this
correspondence seems to be sensitive to different  initial conditions (not shown) and, in any case, we do not have any understanding of why the residual-energy scale 
should growth should be ${\propto}a^{1/2}$, so we will not emphasize this point further. }
The $\tilde{E}^{+}$ decay, in contrast, matches the simplest prediction of \cref{eq: plus decay law}, with $\sigma_{\theta}\lamp/\lamm\approx1$
and $\tilde{E}^{+}\propto a^{-1}$. This feature seems  robust across different initial conditions with sufficiently high $\chi_{\rm exp0}$ and suggests physically 
that the  dominant scale of $\zm$ that advects $\zp$ to cause dissipation ($\lamm$) is the same 
as that which governs the evolution of $\zm$ ($\lamp$). The reason for such a  correspondence is not immediately obvious but may be related 
to the fact that the scales that control the growth of $\zm$ are also coherent with $\zp$ (being driven by reflection), and thus most effective at advecting 
and dissipating $\zp$. Another, non-exclusive possibility is 
that the $\zm$ that is most effective at advecting $\zp$ has a 
 ${\sim}k_{\perp}^{-3}$ spectrum (as motivated above), which would yield a nonlinear turnover time ($\taup\sim a^{3/2}\lamm/\zm$) that is independent of scale. 
Perhaps also of note is that a self-similar power-law decay is possible in \cref{eq: plus decay law} only if $\sigma_{\theta}\lamp/\lamm$ is constant.

The lower-amplitude simulations, with $\chi_{\rm A0}=0.75$ ($\chi_{\rm exp0}=7.5$; dash-dotted lines) and $\chi_{\rm A0}=0.075$ ($\chi_{\rm exp0}=0.75$; colored dotted lines) behave rather 
differently. The $\chi_{\rm A0}=0.75$ shows a small amount of decay in $\tilde{E}^{+}$, while the $\chi_{\rm A0}=0.075$ case shows almost none, 
and $\tilde{E}^{-}$ grows much more rapidly and is not a power law in either simulations. We will show in \cref{sec: Balanced phase} that 
this behavior is effectively just the linear growth of  long-wavelength  $\zm$  modes, which are $k_{z}=0$ modes seeded from the initial conditions in the 
simulation. The linear growth of such modes  is significantly faster than the nonlinear prediction \eqref{Eq:Minus_Law}, so the system 
reaches the balanced regime at smaller $a$ (equivalently, the nonlinear prediction is $\zm\sim \zp/\chi_{\rm exp}$ and $\chi_{\rm exp}$ is not large initially).  The lack of $\tilde{E}^{+}$ decay is a consequence of the turbulence
being weak, or,  in the case of the $\chi_{\rm A0}=0.75$ simulation,  rapidly becoming so, because $\chi_{\rm A}\sim (\taum)^{-1}/(k_{\|}\va)\propto a^{-1/2}$ for fixed $\zp$
and $k_{\|}$.  
We have observed generically that weak turbulence in the EBM exhibits almost no nonlinear decay, behaving 
effectively as a collection of linear modes. However, we caution that key aspects of the expanding-box approximation are not valid for modes in the weak regime, 
and its predictions  for how $\bzm$ is forced via randomly-phased $\bzp$ are likely incorrect.\footnote{In particular, in the weak regime, a $\bzm$ fluctuation 
sourced via reflection can, for some parameters, propagate backwards across a distance larger than the box length. In doing so, it will re-encounter the same $\bzp$ fluctuations
that sourced it, thereby introducing artificial correlations. For linear Fourier modes, which are periodic by fiat, this correlation causes a reflected $\bzm$ wave to oscillate
as a standing wave without growing in time. In contrast, Ref.~\cite{Chandran19} argue that $\zm$ could 
build in time via a random walk because such correlations get scrambled, leading to a prediction that is similar to the strong phenomenology \cref{Eq:Minus_Law}.}
Further work is needed to understand these issues, but weak-turbulence EBM results should be treated with caution.

\begin{figure*}
\centering
\includegraphics[width=0.85\textwidth]{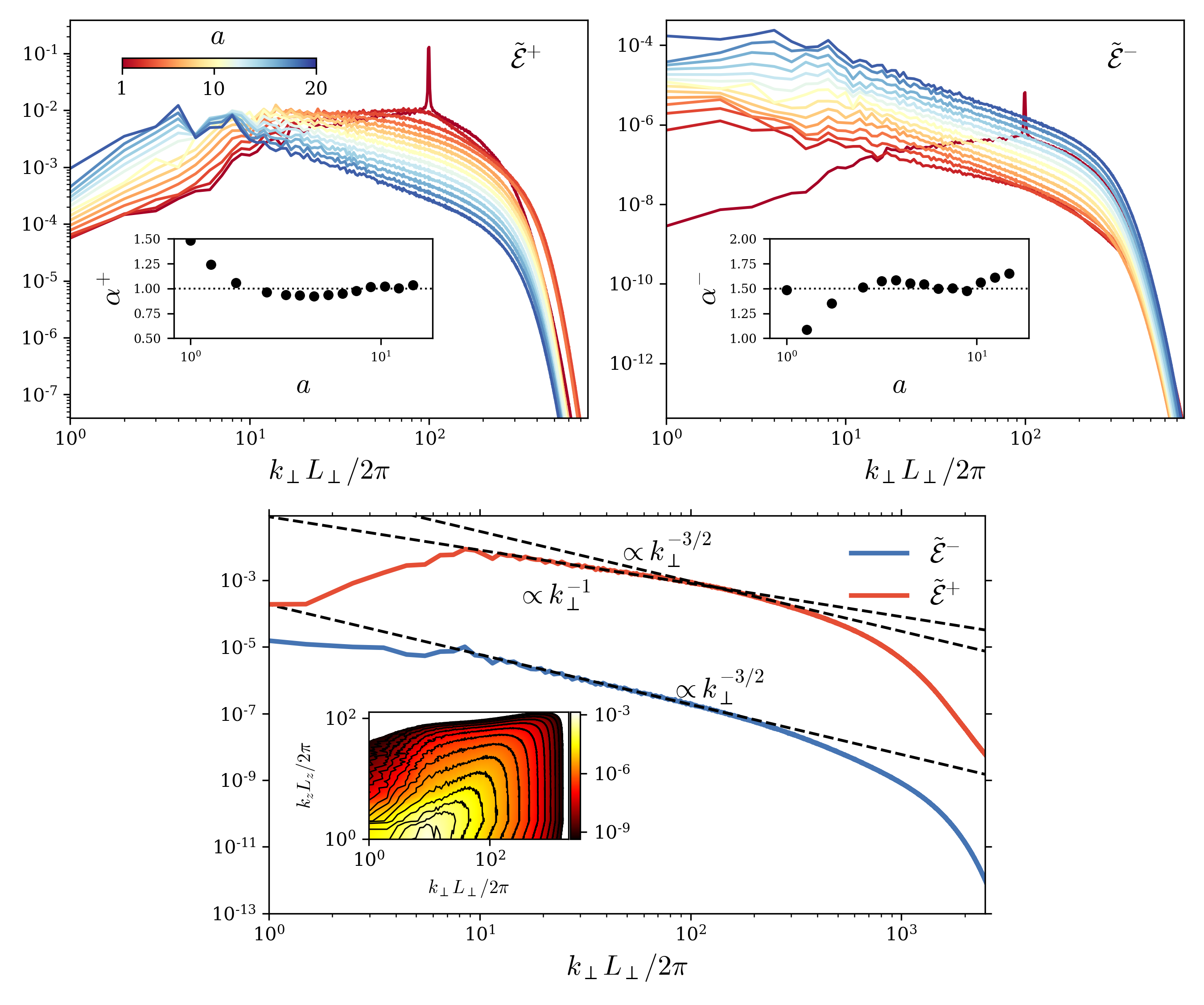}
\caption{ Wave-action energy spectra $\tilde{\mathcal{E}}^{\pm}(k_{\perp})$ during the imbalanced phase of the simulation. 
The top-left and top-right panels show $\tilde{\mathcal{E}}^{+}(k_{\perp})$ and  $\tilde{\mathcal{E}}^{-}(k_{\perp})$, respectively (note the differing vertical scales), 
with the different colors showing different time/radii, as indicated by the color bar. In each panel, the inset shows the best-fit power-law spectral slope, which is fit below 
the measured correlation scale at each $a$. The bottom panel shows both $\tilde{\mathcal{E}}^{+}$ (red) and $\tilde{\mathcal{E}}^{-}$ (blue) at $a\approx5$ when the simulation 
is refined to the higher resolution of $n_{\perp}^{2}\times n_{z} = 8192^{2}\times 256$. Dashed back lines show various power-law slopes, highlighting a steepening of $\mathcal{\tilde{E}}^{+}(k_{\perp})$ at small scales (although there is not sufficient range to say whether it steepens to $\tilde{\mathcal{E}}^{+}\propto k^{-3/2}$ as observed in 
the solar wind). The inset shows the two-dimensional spectrum of the dominant waves $\tilde{\mathcal{E}}^{+}(k_{\perp},k_{z})$, illustrating 
how the fluctuations have decorrelated in the parallel direction (as indicated by the approximately vertical contours at larger $k_{\perp}$).\label{fig: imbalanced spectra}}
\end{figure*}

\subsection{Turbulent spectra}\label{sub: imbalanced spectra}

The energy spectra $\tilde{\mathcal{E}}^{\pm}(k_{\perp})$ for the $\chi_{\rm A0}=96$ ($\chi_{\rm exp0}=960$) simulation over this imbalanced phase are shown in \cref{fig: imbalanced spectra}. The 
two top panels show the time evolution of $\tilde{\mathcal{E}}^{+}$ and $\tilde{\mathcal{E}}^{-}$, respectively, demonstrating their very different evolution. The bottom panel
shows the simulation at $a=5.2$ when it has been refined to a resolution $n_{\perp}^{2}\times n_{z}=8192^{2}\times 256$ in order to attempt to capture the 
transition to standard imbalanced turbulence at small scales. 
The obvious feature of $\tilde{\mathcal{E}}^{+}(k_{\perp})$ is its rapid migration towards large scales, which will be discussed in detail below in  \cref{sub: anastrophy}. As this occurs, 
$\tilde{\mathcal{E}}^{+}$ develops a wide $\tilde{\mathcal{E}}^{+}\propto k_{\perp}^{-1}$ range, which eventually transitions into a steeper slope at small scales (see lower
panel at $a\approx 5.2$). While the simulation does not have sufficient resolution to easily diagnose the slope of this smaller-scale turbulence, 
it is consistent with $\tilde{\mathcal{E}}^{+}\propto k_{\perp}^{-3/2}$, as would be expected at small scales once nonlinear shearing rates inevitably overwhelm reflection-related
physics (see below). The  evolution of  $\tilde{\mathcal{E}}^{-}(k_{\perp})$ is quite different, rapidly moving to large scales  at very early times. 
This feature is consistent with the discussion above, where we argued  that the dominant scale of $\zm$ has no reason to match that of $\zp$, because
large amplitude $\zp$ eddies cause both stronger growth and stronger damping. The
spectral slope rapidly reaches $\tilde{\mathcal{E}}^{-}\propto k_{\perp}^{-3/2}$ over wide range of scales that overlaps with the range where $\tilde{\mathcal{E}}^{+}\propto k_{\perp}^{-1}$.

\begin{figure*}
\centering
\includegraphics[width=1\linewidth]{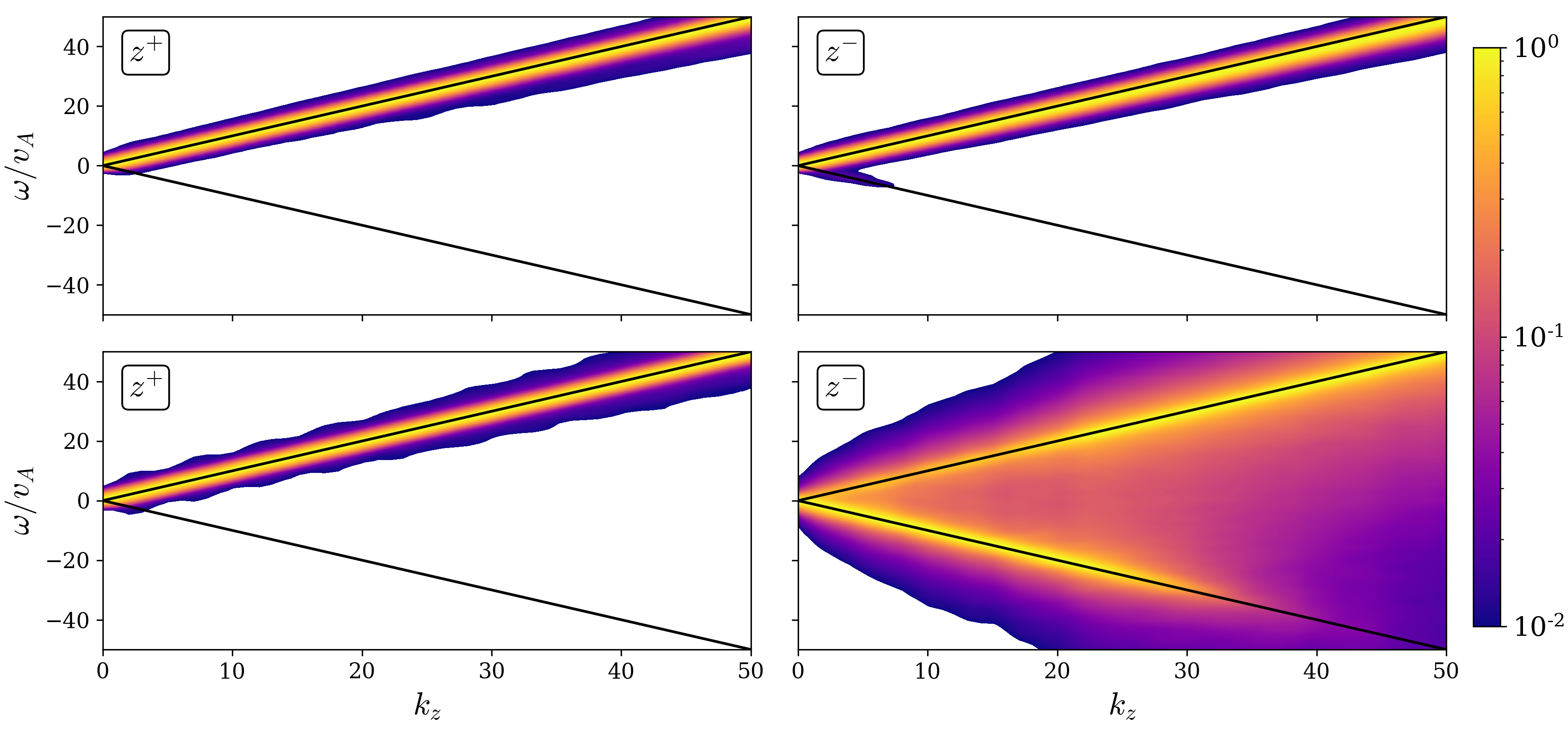}
\caption{ Space-time Fourier transform \cref{eq: spectrogram def} of $\bzp$ (left panels) and $\bzm$ (right panels). Each column 
is normalized to its maximum value to better illustrate the structure. The top panels show the $\chi_{\rm exp0}=960$ reflection-driven turbulence 
simulation at $a\approx 5$ (as in \cref{Fig:Snapshot}); the bottom panels show the same simulation around the same time, but restarted with the 
reflection and expansion terms artificially removed (\textit{viz.,} as a normal decaying RMHD turbulence starting with initial conditions generated from the 
reflection-driven turbulence). While $\bzm$ fluctuations remain anomalously coherent with $\bzp$ in the reflection-driven simulation (top panels), the homogenous 
decaying turbulence does not exhibit this feature (the dominance of outwards-propagating $\zm$ fluctuations at $k_{z}\gtrsim 20$ in the 
bottom-right panel is likely due to field-line wandering and the diagnostic should not be trusted in this range).\\
}
\label{Fig: spectrogram}
\end{figure*}

These basic features  can be plausibly understood within the framework discussed above if we also consider that $\zm$ could 
consist of two qualitatively separate  ``classical'' and ``anomalous'' components, as introduced in previous works \cite{Velli89,Verdini09,Perez13,Chandran19}.
The anomalous component maintains coherence with $\zp$, allowing it to shear coherently over long times 
and thus dominating $\zp$'s turbulent decay. The classical part, in contrast, would be that cascaded from larger scales in 
$\zm$, dominating the measured spectrum but only weakly affecting the decay of $\zp$ because the nonlinear
interactions are weak and accumulate as a random walk. Indeed, the claim above --- that $\zm$ should form a $\tilde{\mathcal{E}}^{-}\propto k_{\perp}^{-3}$ spectrum due to 
the balance between reflections and nonlinearity --- is
 not sustainable towards small scales. In particular, 
in order to form a $k_\perp^{-3}$ spectrum, the energy  injection at each scale from reflection must be larger than the flux arriving to this scale from larger scales due to nonlinear transfer. Based on the phenomenology of \cref{sub: dmitruk decay} and using $\tilde{\mathcal{E}}^{+}\propto k_{\perp}^{-1}$, one finds that the injected flux scales as $\varepsilon^-\propto k_\perp \zp(\zm)^2\propto k_\perp^{-1}$, implying that it declines towards smaller scales and will be overwhelmed by the nonlinear transfer from larger scales \cite{Verdini09}.
This idea can thus be used to motivate there being a ``hidden'' $\tilde{\mathcal{E}}^{-}\propto k_{\perp}^{-3}$ spectrum  in \cref{fig: imbalanced spectra} that 
is the dominant advector of the $\zp$ (interestingly, the measured spectrum of the 2-D modes does follow $\tilde{\mathcal{E}}^{-}\propto k_{\perp}^{-3}$; not shown). As noted by Ref.~\cite{Velli89}
and extended to more realistic anisotropic turbulence by Ref.~\cite{Perez13}, because the $\zp$ cascade rate is $\varepsilon^{+}\propto k_{\perp}\zm (\zp)^{2}$, 
if this is independent of $k_{\perp}$ (a constant-flux $\zp$ cascade), the $\zp$ spectrum would be $\tilde{\mathcal{E}}^{+}\propto k_{\perp}^{-1}$ as observed here, in previous reflection-turbulence simulations \cite{Verdini09,Perez13,Chandran19,Squire2020},  and in the solar wind.

\subsection{Anomalous coherence}\label{sub: anomalous coherence}

The ``coherence assumption''  was used extensively in the discussion above in order to justify using the
nonlinear time $\taup\sim a^{3/2}\lamm/\zm$ to estimate the turbulent decay rate of the $\zp$ fluctuations, even though the 
$\zm$ fluctuations are very low amplitude and thus might be expected to cascade $\zp$ weakly.  In \cref{Fig: spectrogram},
we diagnose this assumption numerically using space-time Fourier spectrum \cite{Meyrand12,Lugones19}, defined as
\begin{equation}
\tilde{\mathcal{E}}^{\pm}(k_z,\omega)=\dfrac{1}{2} \left\langle\vert \hat{\bm{\tilde{z}}}^{\pm}(k_z,\omega)\vert^2\right\rangle_\perp,\label{eq: spectrogram def}
\end{equation}
where $\hat{\bm{\tilde{z}}}^{\pm}(k_z,\omega)$ are the Fourier transforms in time and space of the Els\"asser field. The average, $\left\langle 
\cdot \right\rangle_\perp$, is taken over all perpendicular wavenumbers, meaning that $\tilde{\mathcal{E}}^{\pm}(k_z,\omega)$ will be dominated 
by contributions from the perpendicular scales that dominate the energy spectrum at each $k_z$. 
In the absence of reflection, linear $\zpm$ perturbations satisfy the dispersion relation $\omega^{\pm}=\pm  k_{z} \va $, so
would each show up as a single line in $\tilde{\mathcal{E}}^{\pm}(k_z,\omega)$ at $\omega = \pm k_{z} \va$ (we take $k_{z}>0$). 
In the nonlinear simulation, the $\omega$ location of the peak of  $\tilde{\mathcal{E}}^{\pm}(k_z,\omega)$ versus $k_{z}$ thus indicates the effective 
velocity of $\zpm$ perturbations, while its width provides a measure of the  the level of nonlinear broadening due to the turbulence. 
Note that, because the Fourier transform is taken in $k_{z}$, rather than $k_{\|}$, care is required to ensure that the diagnostic is not affected by field-line wandering. We 
will see  that this likely pollutes the results for $k_{z}\gtrsim 25$ in our simulations.

In the top panels of \cref{Fig: spectrogram}, we show $\tilde{\mathcal{E}}^{+}(k_z,\omega)$ (left) and $\tilde{\mathcal{E}}^{-}(k_z,\omega)$ (right)
 in the $\chi_{\rm A0}=96$ reflection-driven simulation. It is normalized to its maximal value at each $k_{z}$ and computed over several Alfv\'en crossing times around
    $a\approx 5$. As expected, the $\zp$ fluctuations concentrate in the vicinity of the Alfv\'en-wave prediction\footnote{Alfv\'en-wave frequencies are reduced 
    slightly by expansion (see \cref{sub: linear balanced}), but the effect
    is negligible for the range plotted here.} $\omega \approx k_{z}\va$,
 with modest nonlinear broadening. But, the $\zm$ fluctuations (top-right panel) are seen to propagate oppositely to linear Alfv\'en waves, populating the same (upper) region as the $\zp$. This provides direct empirical evidence that they propagate together with $\zp$, leading to anomalous coherence. 
In the frame of the $\zp$ fluctuations, such $\zm$ are stationary, and can thus coherently shear the $\zp$ eddy over the timescale $\taum$.

To assess the role of reflection in supporting this phenomenon, in the
bottom panels we illustrate the same plots, but for standard homogenous decaying turbulence. Specifically, we restart the reflection-driven simulation from the same time
pictured in the top panels, but with the reflection and expansion terms removed, then allow this turbulence to decay for several Alfv\'en time to 
measure $\tilde{\mathcal{E}}^{\pm}(k_z,\omega)$ (over this timeframe, $\zm$ decays noticeably, but $\zp$ does not, meaning the effect of $\zp$ on $\zm$ should remain similar). While $\tilde{\mathcal{E}}^{+}(k_z,\omega)$ (bottom-left panel) remains similar, we see a much wider spread in $\tilde{\mathcal{E}}^{-}(k_z,\omega)$ (bottom-right panel), which extends down 
to  $\omega \approx -k_{z} \va$. These general features are as expected because the $\zp$ modes  shear the $\zm$ modes with a nonlinear time comparable to their linear time, thus forming a nonlinear frequency spread of width ${\sim }k_{z}\va$. 
The change to $\omega>0$ dominating  around  $k_{z}\gtrsim 25$ is artificial, occurring because our Fourier transform in $k_{z}$ does not correctly follow the field 
lines, causing the measurement to be dominated by the advection of high-$k_{\perp}$  structures (presumably this same effect occurs in the 
left panels also, but is hidden because the fluctuations already sit at $\omega>0$). 

The simplest way to understand these results is as a direct numerical demonstration of
the importance of reflection in maintaining anomalous coherence in imbalanced turbulence  \cite{Chandran19}.   The top panels of \Cref{Fig: spectrogram} verify
that the $\zm$ effectively remain stationary in the frame of  $\zp$ fluctuations; they thus do not undergo 
Alfv\'en-wave collisions and can shear $\zp$ coherently to enable a strong cascade.
While similar  ideas have appeared in a number of previous works for both homogenous and reflection-driven turbulence \cite{Velli89,Lithwick07,Perez13,Chandran19,Schekochihin2022},
 our results here provide a particularly clear demonstration of the effect and, via the 
comparison of the top- and bottom-right panels in \cref{Fig: spectrogram}, establish the importance of reflection in maintaining the coherence. Interestingly, Ref.~\cite{Lugones19}
have reported similar, though less extreme, behavior of $\zm$ in homogenous imbalanced MHD turbulence simulations with external forcing. 
While this does not directly disagree with our results here (since the bottom panels in \cref{Fig: spectrogram} are decaying), the topic clearly 
deserves more study to understand the impact of forcing (via reflection or otherwise) on coherence.

\subsection{Wave-action anastrophy growth and the split cascade}\label{sub: anastrophy}

\begin{figure}
\centering
\includegraphics[width=1.0\linewidth]{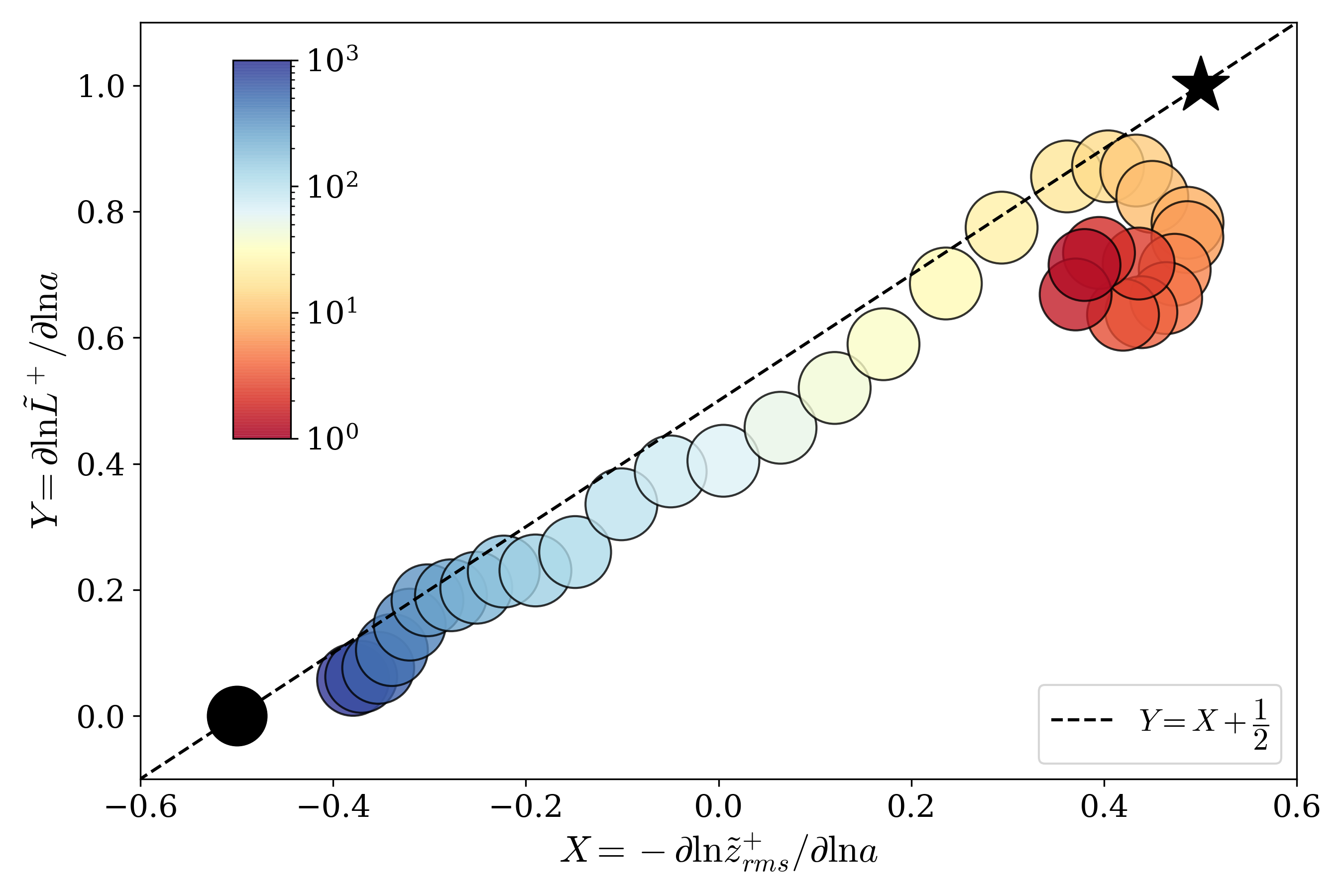}
\caption{Parametric representation of the instantaneous scaling exponents of $1/\zp_{\text{rms}}$ and the energy correlation length $\Lp$  during 
the radial transport. The colors  indicate the normalized radial distance  $a$ (in logarithmic space). The dashed line $Y=X+1/2$ represents the 
theoretical expectation based on anomalous growth of anastrophy 
(\cref{Eq: L+ growth prediction}). The black star corresponds to the expected position for an 
anastrophy-conserving decay characterized by $\tilde{E}\propto a^{-1}$, as described in \cref{sub: dmitruk decay}. The black dot corresponds to the 
asymptotic expectation based on the linear solution (\cref{sub: linear balanced}) for the long-wavelength expansion-dominated  modes with $\Delta <1/2$, which 
dominate the simulation at late times.}
\label{Fig:XY}
\end{figure}

In this section we argue that the turbulent growth of   ``wave-action {anastrophy}''  (wave-action magnetic vector potential squared) causes $\Lp$, the co-moving correlation scale of $\zp$, to rush to large scales as $\zp$ decays. This effect places a strong constraint
on the nonlinear dynamics with interesting implications for the solar wind. 
It can be equivalently viewed in the expanding (physical) frame  as the turbulent suppression of 
 anastrophy decay  compared to what occurs for linear waves.

\subsubsection{Wave-action anastrophy}

Our starting 
point is to note that, because
 $\gradp\cdot\bzpm=0$, $\gradp\cdot\bb=0$, and $\gradp\cdot\bu=0$, one can define the wave-action potentials:
\begin{equation}
    \zh\times \gradp\zetapm = \bzpm,\: \zh\times\gradp\az = \bb,\: \zh \times \gradp\phit = \bu.\label{Eq: potential defs}
\end{equation}
Here, $\gradp$ is the co-moving-frame gradient, so these potentials differ from those naturally defined in the physical (expanding) frame, but will be more convenient here.\footnote{Accounting for the 
various factors of $a$ in gradients and the Alfv\'enic normalization of $\bb$, one finds that $\az$ is related 
to the physical vector potential $\egrad\times \bm{A}=\bm{B}$ by $\az = a^{1/2} A_z$.}
Equation~\eqref{Eq:Equations} can then equivalently be written in terms of $\zetapm$, or   $\phit$ and $\az$, which evolves as
   \begin{align}
\dot{a}\frac{\partial \az}{\partial \ln a}&+\frac{1}{a^{1/2}}\{\phit, \az\}  =\vao \frac{\partial \phit}{\partial z} + \frac{\dot{a}}{2} \az, \label{eq: Az equation}
\end{align} 
where the Poisson bracket is defined as $\{  \phit,\az \} = {\hat{\mathbf{z}}}\cdot \gradp \phit\times \gradp \az$ \cite{Schekochihin2009}.
Multiplying \eqref{eq: Az equation} by $\az$ and integrating, we form the equation for \textit{wave-action anastrophy}, $\anas\equiv \langle \az^2\rangle/2$:
  \begin{equation}
\dot{a}\left(\frac{\partial \anas}{\partial \ln a} - \anas\right) =  \vao \left\langle \az \frac{\partial \phit}{\partial z}\right\rangle = \frac{\vao}{2} \left\langle \zetap \frac{\partial \zetam}{\partial z}\right\rangle.\label{Eq: anastrophy conservation}
\end{equation}  
The nonlinear term has disappeared because anastrophy is an ideal invariant of the 2D RMHD system, while the expansion causes  $\anas$ to grow (the $-\anas$ on the left-hand-side of \eqref{Eq: anastrophy conservation}) and the 3-D term  $\langle \zetap\partial_z\zetam \rangle = - \langle \zetam\partial_z\zetap \rangle$ 
can in principle either destroy or create it, depending on the correlation between
the two Els\"asser fields. Omitted in \cref{Eq: anastrophy conservation} is an additional hyper-dissipation term on its right-hand side, 
which can dissipate small-scale $\anas$ and thus provide an important contribution if there exists a turbulent
flux of $\anas$ to small scales.

\Cref{Eq: anastrophy conservation} shows that if $\langle \zetap\partial_z\zetam \rangle$ is small in the appropriate sense,
wave-action anastrophy will \textit{grow} rapidly (up to $\anas\propto a$), 
purely due to linear expansion effects.\footnote{Note that in physical variables, this scaling $\anas\propto a$ corresponds to $A_z$ itself
being constant with $a$, so that the anastrophy  $\mathcal{A} = \int d{V}\,A_z^2$ scales as $\mathcal{A}\propto a^2$ (the physical  volume of integration $d{V}$ increases ${\propto }a^2$); this is a consequence of the fact that at very low frequencies, $\bm{B}_\perp\propto a^{-1}$ due to flux conservation, while perpendicular lengthscales increase ${\propto}a$.} As a  relevant example, if the fluctuations satisfy $|\sigma_{\theta}|=1$ ($\zetap\propto \zetam$), lying on the edge of the circle 
plot in the right panel of \cref{Fig: time evolutions}, then $\langle \zetap\partial_z\zetam \rangle =0$,  driving growth of $\anas$. 
We will now argue that in strong reflection-driven turbulence,  the wave-action anastrophy grows with $a$, 
even in 3-D.
The argument relies on considering what occurs for propagating linear Alfv\'en waves, which, so long as $\Delta=k_{z}\vao/\dot{a}>1/2$ (see \cref{sub: linear balanced}), propagate with constant amplitude on average, and 
thus constant $\anas$. This implies that $\langle \zetap\partial_z\zetam \rangle$ in Eq.~\eqref{Eq: anastrophy conservation}
must exactly  balance the expansion-induced growth. Indeed, as shown in \cref{app: linear anastrophy}, as
an outwards ($\zetap$) fluctuation propagates, the reflected $\zetam$ component trails it by $\pi/2$ in phase and has exactly the required amplitude to ensure that $\vao\langle \zetap\partial_z\zetam \rangle = -2\dot{a} \anas$. Because the phase offset of $\pi/2$ causes  $\langle \zetap\partial_z\zetam \rangle$ to be as negative as possible,
this implies that so long as $|\zetam|/|\zetap|$ remains similar to (or less than) the linear solution, 
\textit{any} change to the phase offset between $\zetam$ and $\zetap$ will increase $\langle \zetap\partial_z\zetam \rangle$ (decrease $|\langle \zetap\partial_z\zetam \rangle|$), thus causing $\anas$ to grow with $a$.

For application in strong reflection-driven turbulence, it is therefore helpful to compare  $\zm$ in the phenomenology of \cref{sub: dmitruk decay} to what the 
the linearly reflected $\zm$ would be for a given $\zp$, knowing that, if its phase offset is perfect, the latter destroys $\anas$ at  just the correct rate to maintain constant $\anas$.
The nonlinear phenomenology yields $\zm\sim\zp/\chi_{\rm exp}$ (\cref{sub: dmitruk decay}), while the  linearly reflected component 
is $\zm \sim \zp/\Delta$ (see \cref{app: linear anastrophy}).
Therefore, the ratio of the two is the critical balance parameter $\chi_{\rm A}$ --- a sensible expectation given that $\chi_{\rm A}$ is the ratio of the two 
effects (Alfv\'enic propagation and nonlinearity) that can compete with expansion to halt the growth of $\zm$.
This implies that in strong ($\chi_{\rm A}\sim 1$) reflection-driven turbulence, the amplitude of the growing $\zm$ is no larger than the 
amplitude needed to maintain constant $\anas$. The consequence is that \textit{any} modification to the linear ($\pi/2$) phase offset between 
$\zm$ and $\zp$ will decrease $\vao|\langle \zetap\partial_z\zetam \rangle|$ below $2\dot{a}\anas$, thereby causing $\anas$ to grow.
While chaotic nonlinear interactions will generically act to scramble the phases of $\zetapm$, we argue that reflection turbulence causes a more pronounced effect: the 
 anomalous coherence, which leads   to the high observed correlation  between  $-\bzm$ and $\bzp$ (negative $\sigma_\theta$), also precludes a large correlation between $\zetap$ and $\partial_z\zetam$.\footnote{For  individual Fourier modes, $\zetapm_{\bm{k}}$,
this follows from the fact that $\langle \zetap\partial_z\zetam \rangle = -2 k_z {\rm Im}[\zetap_{\bm{k}}(\zetam_{\bm{k}})^* ]$, 
while $\sigma_\theta  = -2{\rm Re}[\zetap_{\bm{k}}(\zetam_{\bm{k}})^* ]/(|\zetap_{\bm{k}}||\zetam_{\bm{k}}|)$. Since ${\rm Im}(z)^2+ {\rm Re}(z)^2= |z|^2$, a large $\sigma_\theta$ (proportionally large ${\rm Re}[\zetap_{\bm{k}}(\zetam_{\bm{k}})^\star ]$) precludes the possibility of $\langle \zetap\partial_z\zetam \rangle$ being large compared to $k_z|\zetap_{\bm{k}}||\zetam_{\bm{k}}|$. For a system with a range of modes, a similar argument
can be made via the Cauchy-Schwarz inequality. }
In other words, the phases are partially scrambled by the turbulence, but with a tendency for correlation between $\zetap$ and $-\zetam$, rather than $\zetap$ and $\partial_{z}\zetam$.
The surprising consequence is that, while the decay rate of wave-action energy increases (up to $\tilde{E}\propto a^{-1}$)
as the turbulence becomes stronger (see \cref{Fig: time evolutions}), the opposite is true of the wave-action anastrophy: it is approximately constant in weak turbulence (where $\langle \zetap\partial_z\zetam \rangle$ remains similar to its linear value), but 
\textit{grows} in strong turbulence.

\subsubsection{The growth of $\Lp$}

From here, the arguments are standard \cite{Kolmogorov41}. 
The wave-action energy, which is almost a true inviscid invariant during the imbalanced phase when $|\tilde{E}^{r}|\ll \tilde{E}$, decays nonlinearly due to the turbulent flux between the co-moving correlation scale $\Lp$ and 
the dissipation scales. But, because  the small-scale dissipation of $\anas$ is proportional to the magnetic energy, 
for small  (hyper-)viscosity, if the nonlinear dissipation of $\tilde{E}$ remains finite, 
the nonlinear dissipation of $\anas$ must be smaller \cite{Montgomery1978,Alexakis18}. Combined with the argument above that
$|\vao\langle \zetap\partial_z\zetam \rangle| \lesssim 2\dot{a} \anas$, we thus expect $\anas$ to grow. 
Then, because $\tilde{E}\sim \anas/\Lp^{2}$ for imbalanced fluctuations, 
if $\tilde{E}$ decays while $\anas$ grows (or even remains constant), 
this leads to remarkable phenomenon: the turbulent decay must progress with $\Lp$ increasing rapidly in time.
Specifically, taking $\anas\propto a$ (assuming $|\vao\langle \zetap\partial_z\zetam \rangle| \ll 2\dot{a} \anas$ and minimal nonlinear dissipation), we find
\begin{equation}
 a^{-1}\tilde{E}^{+} \Lp^2 \sim \text{const.} \implies \Lp 
\propto a,
\label{Eq: L+ growth prediction}
\end{equation} 
where we used $\tilde{E}^{+}\propto a^{-1}$ from \cref{eq: plus decay law}.  This prediction applies to the co-moving frame, implying yet faster increase in scales in the physical frame ($L_{+}\propto a^2$). Note
that this law is more extreme than the standard argument for growing correlation scales in decaying 2-D MHD turbulence, which invokes only the lack of nonlinear decay of anastrophy \cite{Hatori1984,Biskamp1989}. It is also worth clarifying that there is no ``trick'' involved with the wave-action variables here: if 
we were instead to work in physical variables  in the co-moving frame, (co-moving) anastrophy would remain constant, but the energy would  linearly decay ${\propto}a^{-1}$ (and thus turbulently decay ${\propto }a^{-2}$) because $\bm{z}^{\pm}$ naturally decays with $a$.

\begin{figure*}
\centering
\includegraphics[width=0.8\linewidth]{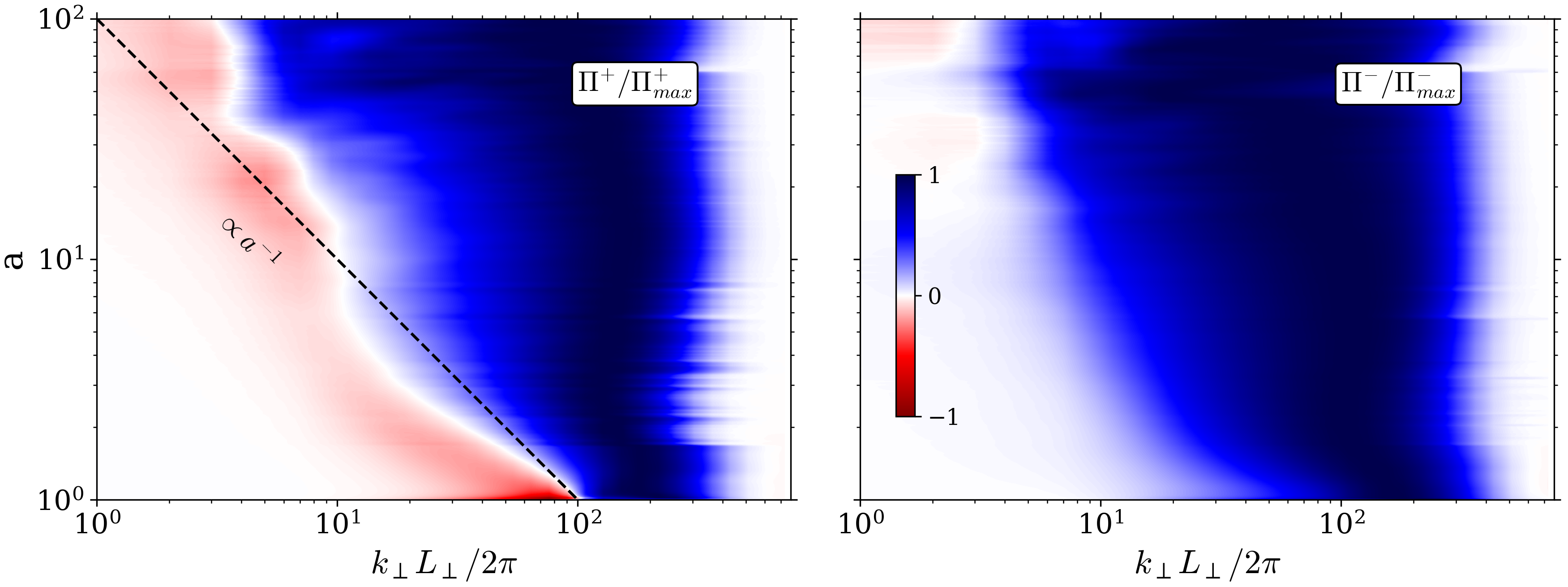}
\caption{Two-dimensional $k_\perp$-$a$ evolution of the Els\"asser fluxes $\Pi^+(k_\perp)$ (left panel) and  $\Pi^-(k_\perp)$ (right panel; see Eq.~\eqref{Eq:Fluxes}). 
At each $a$ the $\Pi^\pm(k_\perp)$ are normalized by their maximum over $k_\perp$ in order to better show their structure. We 
see clear evidence of a split cascade in $\bzp$, with a break between the forward and inverse cascades that migrates to larger scales with time. Although
the cause of the modest deviations from the ${\propto}a^{-1}$ scaling remains unclear, the general behavior is consistent with the discussion in the text and the evolution of the correlation length in \cref{Fig:XY}. }
\label{Fig:Fluxes}
\end{figure*}

The prediction \cref{Eq: L+ growth prediction} is tested in \cref{Fig:XY}. 
We compute the parametric representation,
\begin{equation}
X(a)=-\dfrac{\partial \ln\zp_{\text{rms}}(a)}{\partial \ln a}, \quad Y(a)=\dfrac{\partial \ln\Lp(a)}{\partial \ln a}, 
\end{equation}
where $\zp_{\text{rms}} = \sqrt{2\tilde{E}^{+}}$ and $\Lp$ is computed as
\begin{equation}
   \Lp \equiv \int dk_\perp 
\mathcal{\tilde{E}}^{+}(k_\perp)/k_\perp.
\end{equation}
 $X(a)$ and  $Y(a)$ are the instantaneous scaling exponents of $1/\zp_{\text{rms}}$ and $\Lp$, 
implying that if wave-action anastrophy,  $\anas\sim \tilde{E}^{+}\Lp^{2}$, grows as $\anas\propto a$ during the decay, then
\begin{equation}
Y(a)=X(a)+\dfrac{1}{2}.\label{eq: xy anastrophy pred}
\end{equation}
This relation is independent of the decay rate of $\tilde{E}$ and thus the decay phenomenology. 
We see in \cref{Fig:XY} that all through the imbalanced phase ($a\lesssim 50$), 
$X$ and $Y$ sit almost on the line \eqref{eq: xy anastrophy pred}, implying 
$\Lp$ grows almost as predicted by wave-action anastrophy growth (slightly more slowly). In the later dynamics, 
which will be described in more detail below, the fluctuation decay/growth rate ($X$) changes significantly, 
but wave-action anastrophy remains ${\propto}a$ as indicated by its evolution along the dotted line.

\subsubsection{The split cascade}

Physically, the fast increase in $\Lp$ implies the energy  decays through a split cascade, whereby it is  forced to flow to both  small and large scales simultaneously.
We diagnose this surprising phenomenon directly
in \cref{Fig:Fluxes} by
 computing the Els\"asser perpendicular 
wave-action-energy fluxes as a function of perpendicular wavenumber $k_\perp$ \cite{Frish95}:
\begin{equation}
\Pi^{\pm}(k_\perp)=-a^{-3/2}\dfrac{2\pi}{L_{\perp}}\int \dfrac{d^3\bm{r}}{V}\left[\bzpm \right]_{k_\perp}^{<}\cdot(\bzmp\cdot\gradp\bzpm), 
\label{Eq:Fluxes}
\end{equation}
where the low-pass filter is defined by 
\begin{eqnarray}
\left[\bzpm \right]_{k_\perp}^{<}=\sum_{k'_{z}}\sum_{\mid\bm{k}'_\perp\mid\leq k_\perp}e^{i\bm{k}'\cdot\bm{r}} \bzpm_{\bm{k}}.
\end{eqnarray}
The split cascade of the energetically dominant field $\zp$ is revealed by the break between 
the blue and red bands that extends diagonally upwards. It is located near the measured  $1/\Lp$ at earlier 
times, decreasing  as expected   due to the conservation of anastrophy (approximately ${\propto}1/a$). 
On the right of the break, $\tilde{E}^{+}$ cascades towards small scales where reflection 
becomes subdominant and the hyper-viscosity allows its dissipation; on the 
left, $\tilde{E}^{+}$ cascades towards large scales, allowing $\Lp$ to increase in time. 
The break scale deviates modestly from the ${\propto}a^{-1}$ expectation, increasing more rapidly at early times and then slowing somewhat around $a\approx5$ for unknown reasons, but its behavior is broadly consistent with the evolution of $\Lp$ (\cref{Fig:XY}).
  The sub-dominant field $\zm$ undergoes a direct cascade during  
its entire evolution, aside from at the largest scales at late times, where the dynamics start becoming effectively two-dimensional and balanced, differing significantly from the imbalanced phase (see below). 
This leads to the interesting phenomenon whereby  $\zm$ and $\zp$ cascade in opposite directions across a modest range of intermediate scales (those above the break scale in $\Pi^+$) during the imbalanced turbulent decay.
Similar dual, counter-directional Els\"asser cascades have  been reported previously in flux tube simulations of coronal holes
\cite{Ballegooijen17},  and observed in high cross-helicity solar-wind streams \cite{Smith09, Coburn15}.

\section{Balanced, magnetically dominated phase}\label{sec: Balanced phase}

With $\zp$ decaying while $\zm$ grows, it is clear that the 
imbalanced phase must  inevitably end as the fluctuations approach the balanced regime with $\zm\sim \zp$.
Indeed, recall that  the phenomenology of \cref{sub: dmitruk decay} 
predicted  $\zm\sim \zp/\chi_{\rm exp}$, where $\chi_{\rm exp}$ is the ratio of expansion to nonlinear times (\cref{eq: chi exp def}), which is necessarily a decreasing function of time. Thus $\chi_{\rm exp}\sim 1$ marks the end of the imbalanced phase. 
In \cref{Fig: time evolutions}, we saw that the wave-action energy starts \textit{growing} in time, with $\tilde{E}\propto a$, magnetically dominated fluctuations ($\sigma_{r}<0$), and very little turbulent dissipation into 
heat. It is the purpose of this section to understand the important  properties of this balanced, magnetically dominated phase, making predictions for \textit{in-situ} observations at large distances from the sun.

\subsection{Linear EBM dynamics}\label{sub: linear balanced}

\begin{figure}
\centering
\includegraphics[width=1.0\linewidth]{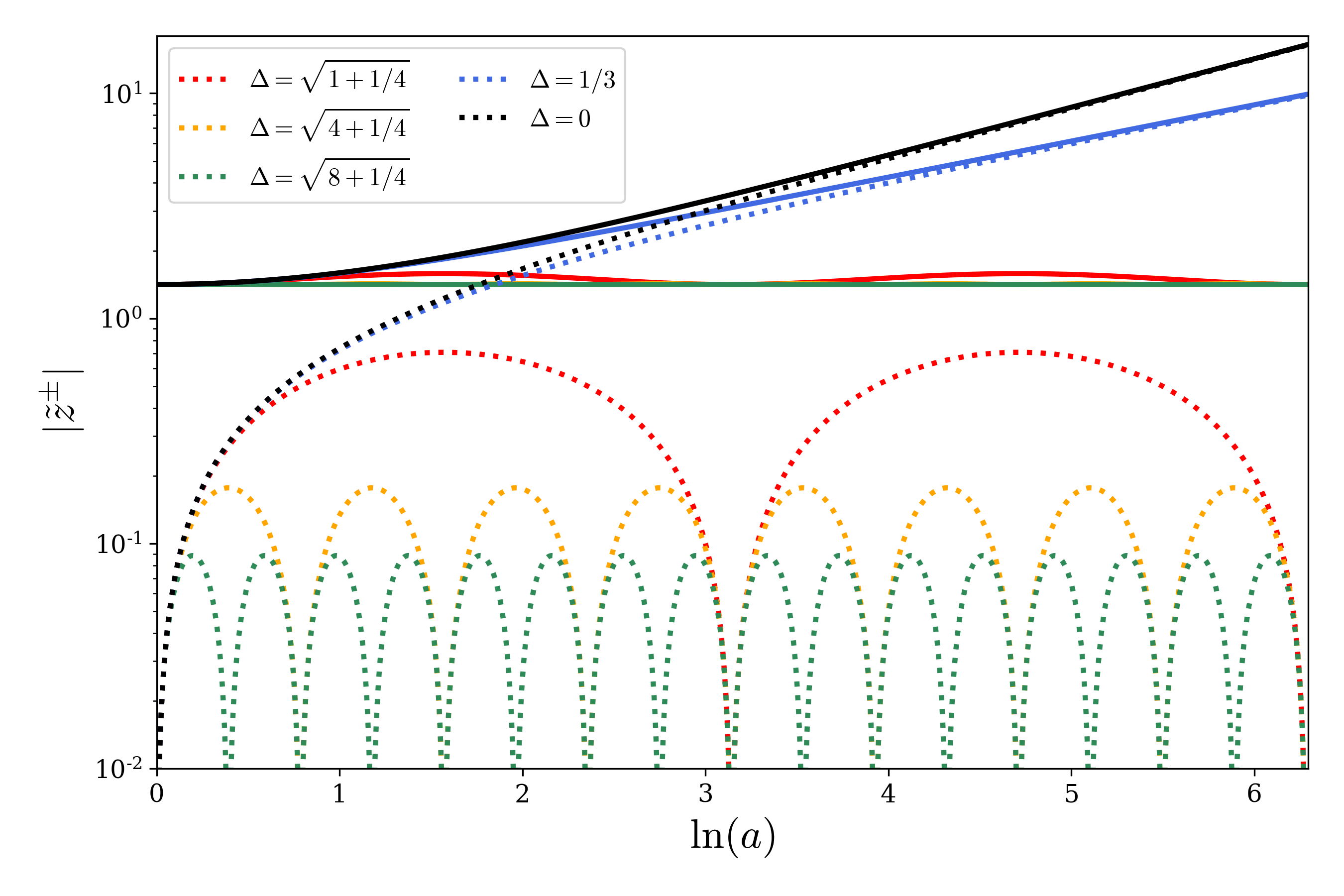}
\caption{Solutions of the linearised equations \cref{eq: linear equations}, starting
from the initial condition $\zm(0)=0$ and $\zp(0)=\sqrt{2}$ with different values of $\Delta$ as labelled. Solid lines show $|\zp(a)|$; dotted lines show $|\zm(a)|$.  $\Delta>1/2$ modes (red, yellow, and green curves),
which are dominated by Alfv\'enic forces, exhibit wave-like behavior with no long-term growth or decay (${\rm Im}(\omega)=0$);  $\zp$ propagates Alfv\'enically with an oscillating phase and approximately constant amplitude, while the amplitude of $\zm$ alternates up and down 
over the wave period as $\zm$ moves in and out of phase with the reflection forcing from $\zp$ (its maximum amplitude scales ${\propto}\Delta^{-1}$; see \cref{Eq:Xipm} and \cref{app: linear anastrophy}).  
In contrast, long-wavelength $\Delta<1/2$ modes  with ${\rm Re}(\omega)=0$ (blue and black curves), do not oscillate like waves at all  because the reflection 
overwhelms the Alfv\'enic restoring force (see \cref{Eq:Eigenfreqs} \cite{heinemann80}). The amplitude of
the magnetically dominated mode grows as $|\zpm(a)|\propto a^{|\omega^\pm|}$, with 
the growth rate $|\omega^\pm|=\tfrac12\sqrt{1-4\Delta^2}$ depending only weakly on $\Delta$ (cf. blue and black curves).}
\label{Fig:Linfig}
\end{figure}

We will show below that by organizing itself into structures that minimize the nonlinear stresses, 
the system becomes effectively linear in its late stages. We thus describe basic features of the linear solution 
here, focusing on the difference between short-wavelength propagating (Alfv\'enic) waves and 
 expansion-dominated solutions at long wavelength, which grow continuously with $a$. These linear solutions are illustrated  in \cref{Fig:Linfig}, starting
 from pure $\bzp$ fluctuations in the initial conditions.   Their characteristics, including the growth of expansion-dominated modes, have been 
studied using various methods in global geometries in a number of previous works \cite{heinemann80,Hollweg90,Hollweg2007}; they are 
not an artefact of the expanding box model.

The full linear solution is 
easily obtained by ignoring the nonlinear terms in \cref{Eq:Equations lna} and assuming 
divergence-free plane-wave solutions of the form $\bzpm = \zpm(a) e^{i k_\perp y + i k_z z} \xh$. 
This gives 
 \begin{equation}
\dfrac{\partial \zpm}{\partial \ln{a}}+\begin{pmatrix}
i\Delta & 1/2 \\
1/2 & -i \Delta
\end{pmatrix}\begin{pmatrix} \zp\\\zm\end{pmatrix}=0,\label{eq: linear equations}
\end{equation}
where $\Delta= k_{z}\va/(\dot{a}/a) = k_{z}\vao/\dot{a}$ (using the time variable $\ln a$ eliminates  explicit time dependence from the linear system), allowing one to insert the ansatz,
\begin{equation}
    \zpm(\ln a) = \zpm_w \exp(i\omega \ln a),
\end{equation}
where $\zpm_w$ is the complex amplitude of $\zpm(a) $. The general solution to \cref{eq: linear equations}
can then be formed via 
 the eigenmodes,
\begin{equation}
\xi^{\pm}=\dfrac{1}{2}\zpm_w \pm i\left(\Delta\mp\omega^{\pm}\right)\zmp_w,
\label{Eq:Xipm}
\end{equation}
which evolve as $\xi^\pm(a) = \xi^\pm_0\exp(i\omega^\pm\ln a)$ from initial conditions $\xi^\pm_0$, where 
the  eigenfrequencies $\omega^\pm$ are 
\begin{equation}
\omega^{\pm}=\pm \sqrt{\Delta^2-1/4}. 
\label{Eq:Eigenfreqs}
\end{equation} 

We see that  $\Delta=1/2$ marks the boundary between oscillating Alfv\'enic modes and growing (or decaying) expansion-dominated modes:  for $\Delta>1/2$, $\omega^{\pm}$ is real and $\zpm$  oscillates with frequency $\omega^\pm$, albeit with a minority-reflected $\zmp$ component that inevitably accompanies 
any $\zpm$ fluctuation; for $\Delta<1/2$, $\omega^{\pm}$ is imaginary and modes grow exponentially,
$\zpm \propto e^{|\omega^\pm|\ln a} = a^{|\omega^\pm|} = a^{\sqrt{1-4\Delta^2}/2}$, because the expansion overwhelms the Alfv\'enic restoring force. 
The growing  expansion-dominated mode, with $\omega = i\sqrt{1/4-\Delta^{2}}$, is magnetically 
dominated with $\bzm\approx-\bzp$ and $|\bb|\gg |\bu|$, while the decaying mode (${\rm Im}(\omega)<0$) is $\bu$ dominated. Physically,
the ${\sim}a^{1/2}$ growth of $\zpm$ corresponds to $|\bm{B}_{\perp}|\propto a^{-1}$ ($|\bm{B}_{\perp}|/|\overline{\bm{B}}|\propto a$) so that $\bm{b}_{\perp}=\bm{B}_{\perp}/\sqrt{4\pi\rho}$ is constant \cite{Velli91,Grappin93}.
Clearly, if there exists any power in such  expansion-dominated modes at early times, they  will inevitably come to dominate the late-time evolution, overtaking the Alfv\'enic ($\Delta>1/2$) modes.

In our simulations with $\Delta_{\rm box}=10$ (\cref{Eq: Delta in sim}), only the $k_{z}=0$ periodic mode 
lies in this expansion-dominated regime. But, the properties of expansion-dominated modes are rather insensitive to 
$k_{z}$ for $\Delta <1/2$: the modes have  no real frequency (oscillating) part and growth rates that exhibit only a small correction compared to the $\Delta=0$ mode (${\rm Im}(\omega^{+})\approx 1/2-\Delta^{2}$ for small $\Delta$). Therefore, we argue that their dynamics should be adequately captured by the simulation, even 
though true $k_{z}=0$ modes are obviously not possible in a realistic non-periodic system.  In reality, 
if we assume that the longest-wavelength modes possible are those of the system scale, with $k_{z}\sim 1/R$, then the 
minimum $\Delta$ available to the system is $\Delta_{\rm min}\sim (\va/R)/(U/R)\sim \va/U<1$.
Thus, in the super-Alfv\'enic ($\va<U$) wind  it is always consistent  to assume that the expansion-dominated modes exist, and indeed,
the range of such modes available to the system will be an increasing function of radius. This feature, whereby $\Delta_{\rm min}$ decreases with radius, 
is clearly not possible to capture in the EBM with a fixed $L_{z}$ (it is captured by global linear solutions \cite{heinemann80,Hollweg2007}), so the impact of this physics should be tested in future work using flux-tube simulations.

\subsection{Transition into the magnetically-dominated phase}

During  the imbalanced phase in our simulations,  the turbulence appears to remain strong with $\chi_{\rm A}\sim 1$, rapidly adjusting its parallel correlation length $\ell_{\|}$ towards larger scales as the turbulence decays (after its initial transient adjustment from the initial conditions, which occurs by $a\approx1.2$).  
This phase ends, and the decay deviates from the $\tilde{E}\propto a^{-1}$ phenomenology of 
 \cref{sub: dmitruk decay}, once it decays sufficiently so that  the box-wavelength modes ($k_{z}=2\pi/L_{z}$) become weak ($\chi_{\rm A}<1$). This  occurs around $a\approx 25$ for the solid lines in \cref{Fig: time evolutions}, which agrees well with the value expected from solving $z^{+}/\lamp\simeq \va 2\pi/L_{z}$ with $\zp\propto a^{-1/2}$ and $\lamp\propto a^{-1/2}$.
 Following this, the expansion-dominated modes inevitably take over, driving the system towards the 
 $|\bb|\gg |\bu|$ linear solution that grows with $|\bb|\propto a^{1/2}$.

 More generally, without the limitations of our periodic box, this transition should be understood by noting that 
if the turbulence remains strong with $\chi_{\rm A}\sim 1$ throughout its decaying imbalanced phase (as appears to be the case until it becomes artificially constrained by the box), 
the transition to the balanced regime, at $\chi_{\rm exp}\sim 1$, will occur when $\Delta =\chi_{\rm exp}/\chi_{\rm A}\sim1 $, \textit{viz.,} at the same time that the dominant modes in the system become expansion dominated. This pleasing consistency of the phenomenology argues that the system cannot 
reach the balanced phase while still dominated by Alfv\'enic physics and suggests that the large scales in the balanced phase will  not be
critically balanced in the usual sense (because their linear physics is dominated by expansion not Alfv\'enic propagation).
This property should hold so long as the turbulence remains strong during the imbalanced-decay phase and transitions into the balanced phase at
$\chi_{\rm exp}\sim1$ (i.e., independently of the evolution of $\lamp$ or other uncertainties in \cref{sub: dmitruk decay})

Following this transition, any further turbulent decay will tend to increase the nonlinear time, thus driving 
the system inevitably towards the linear regime where expansion dominates both  Alfv\'enic and 
nonlinear effects. This can be seen by noting that unless $\lamp$ decreases, then even  the 
fastest-possible linear growth,  $\zp\sim\zm\propto a^{1/2}$, leads to  $\chi_{\rm exp}\sim a^{-1/2} (\zp/\lamp)/\dot{a}$ remaining constant (the dominance of expansion over Alfv\'enic propagation is guaranteed because $\Delta\lesssim1$).
However, we see from the perpendicular structure shown in \cref{Fig:Snapshot} that the system approaches this expansion-dominated state in an interesting and nontrivial way:
rather than simply decaying to low amplitudes to reduce the nonlinear time, it  organizes itself into isolated, coherent structures that approach 
nonlinear solutions in which the magnetic tension balances the pressure. This self organization thus defeats prematurely 
the nonlinear couplings and turbulent dissipation, 
precipitating the system into magnetically dominated ``Alfv\'en vortices'' that  behave almost linearly.

Because the system becomes expansion dominated with little turbulent dissipation, its growth must also satisfy 
the prediction of \cref{eq: xy anastrophy pred} for the growth of wave-action ansatrophy $\anas\propto a$. Thus, during the transition as it moves into the balanced phase, the system evolves downwards along the $Y=X+1/2$ line in \cref{Fig:XY}, tending towards the point $X=-1/2$, $Y=0$ that characterizes purely linear evolution. 


\begin{figure*}
   \includegraphics[width=1\linewidth]{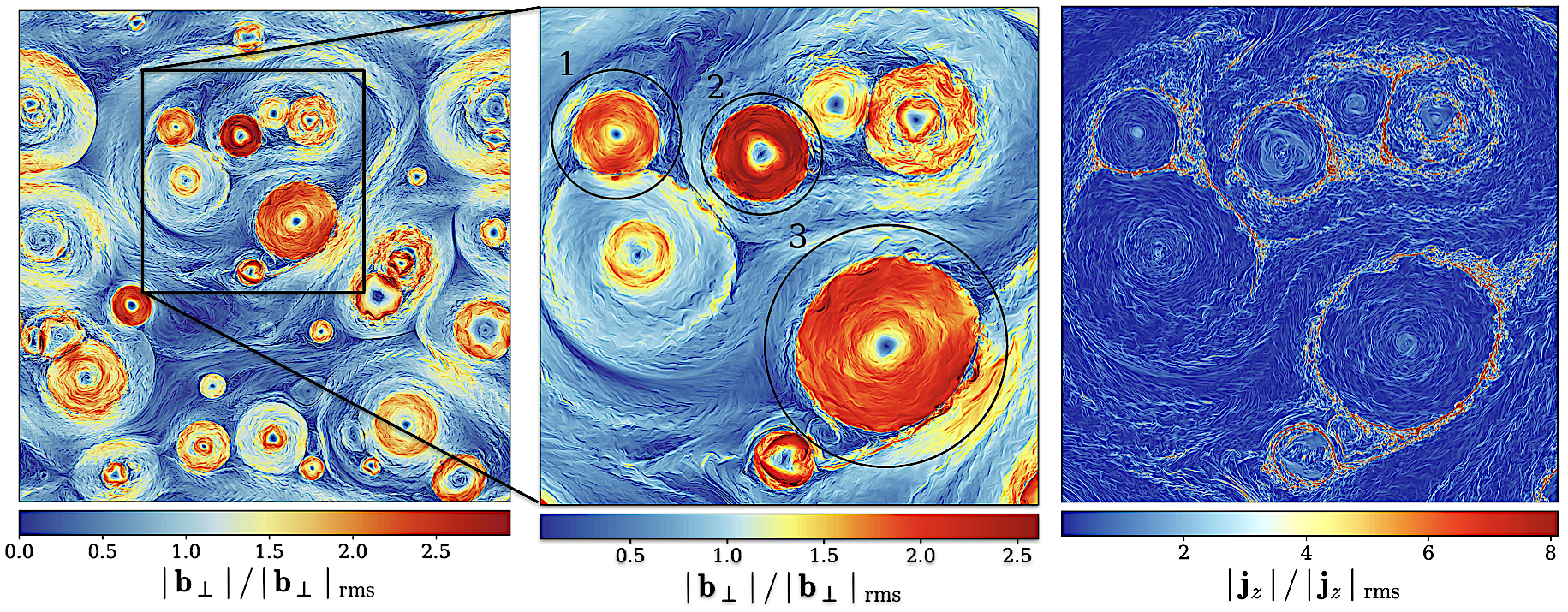}
   \caption{Left: Snapshot of the magnetic field modulus in a plane perpendicular to $B_{0}$ at $a=250$. Middle: Close-up corresponding to the marked region on the left, illustrating Alfv\'en vortices colliding and and merging through reconnection. The black circles mark the regions over which azimuthal averages have been computed to fit the Alfv\'en-vortex solution \eqref{Eq:AV_fit} in figure \ref{Fig:AV_Fit}. Right: Same region as the middle panel, but showing the out-of-plane current. This  reveals sets of intense current rings, a hallmark of the ground-state Alfv\'en vortices.}
   \label{Fig:Alfven_Vortices}
 \end{figure*}
 
  \begin{figure}
\includegraphics[width=1.0\linewidth]{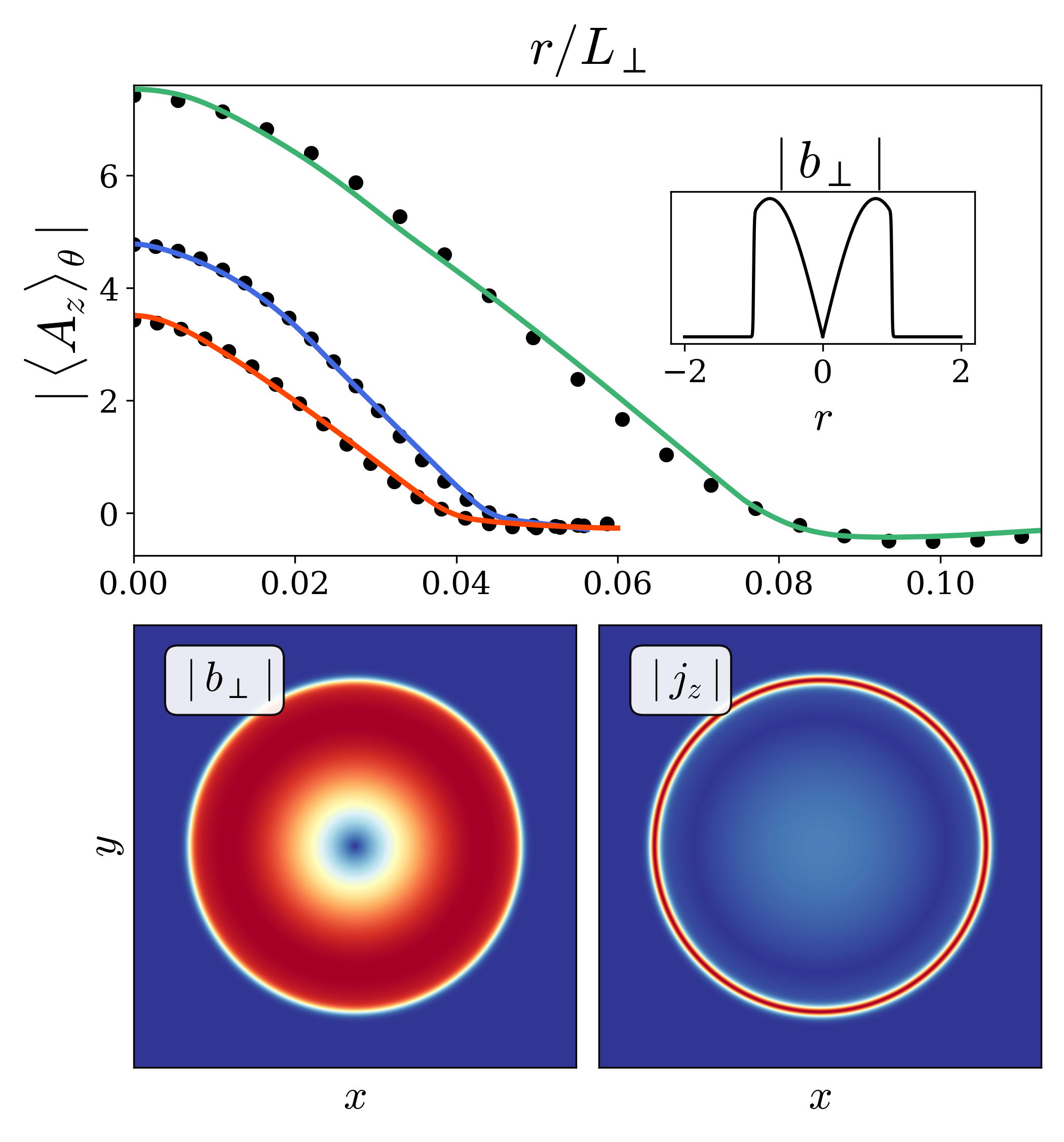}
\caption{Top: Comparison between the absolute value of the magnetic vector potential obtained from numerical simulation at $a=250$ (colored lines) and the analytical prediction \eqref{Eq:AV_fit} (black dots). The red, blue and green lines have been obtained from the Alfv\'en vortices labeled 1,2 and 3 after an azimuthal average about their center (denoted $|\langle\az\rangle_\theta|$). The inset represents the analytical prediction for the magnetic field, highlighting the presence of a discontinuity at the vortex boundary. Bottom: 2D representation of the solution \eqref{Eq:AV_fit} for the magnetic field modulus $| \bm{b}_\perp |$ and absolute value of magnetic current $| 
j_z|$ (the color scales are arbitrary).}
\label{Fig:AV_Fit}
\end{figure}

\subsection{Emergence of Alfv\'en vortices}
\label{Subsec:Vortex}

At this point, the story is mostly finished as far as the turbulent heating and dissipation is concerned:
as the system becomes balanced, it also starts shutting off its nonlinear dissipation, 
creating long-parallel-wavelength perpendicular structures that grow  with $|\bb|\sim a^{1/2}$.
However, the quasi-circular  structures that emerge (see \cref{Fig:Snapshot})  are of significant interest, both for comparison to \textit{in-situ} observations, 
and because they are picturesque illustrations of the ``cellularization'' of turbulence \cite{Matthaeus15} --- a vivid example of self-organization \cite{Horiuchi85}. 
They can be understood using a classical variational argument \cite{Matthaeus80}. Motivated by the turbulent wave-action anastrophy growth, we minimize the Alfv\'enic magnetic energy per unit volume, $\langle |\bm{b}_\perp|^2\rangle/2=a^{-1}\tilde{E}^b,$ subject to the constancy of the anastrophy per unit volume, $\langle A_z^2\rangle/2=a^{-1}\anas$  (by this choice of variables, we factor out the expansion-induced dependence on $a$; both $\langle |\bm{b}_\perp|^2\rangle$ and $\langle A_z^2\rangle$ remain constant under linear evolution for $a\gg1$). Such minimization requires that during the relaxation process the kinetic energy is  dissipated completely, leaving a pure magnetic state. It  is thus aided significantly by  expansion, which
damps $\bm{u}_{\perp}$ but not $\bm{b}_{\perp}$ (see \cref{Eq:induction,Eq:momentum}) and  preferentially increases the energy content of the longest-parallel-wavelength  modes (thus creating quasi-2D dynamics at a large radial distances).

These arguments lead us to the variational problem 
\begin{equation}
    a^{-1}\delta\! \int d^{3}\bm{r}\left( |\gradp \az|^2-\Lambda \az^2\right)=0, 
    \label{Eq:variational}
\end{equation}
where $\delta$ denotes the functional derivative and $\Lambda$ a Lagrangian multiplier. 
Identifying $\Lambda$ with a characteristic scale $K_\perp$ via $\Lambda=-K_{\perp}^2$, the Euler-Lagrange equation becomes the Helmholtz equation,
\begin{equation}
    \gradp^{2}\az=-K_{\perp}^2 \az.
    \label{Eq:Helmholtz}
\end{equation}
Recalling that $\az$ evolves as a passive scalar in 2-D (see Eq.~\eqref{eq: Az equation}), we now imagine some region, or ``cell,'' in the domain that can change shape and mix in order to 
approach the minimum energy state, \textit{viz.,} the solution of \eqref{Eq:Helmholtz} with the minimum possible $K_\perp$. The argument is effectively that 
 the Lagrange multiplier $K_\perp$ should be piecewise constant, enforcing the minimization principle across patch-like ``cells'' where the turbulence becomes suppressed.  We assume the value of $\az$ outside the cell in question 
to be approximately constant (based on  \cref{Fig:Snapshot}, this may not be so unreasonable as it sounds), which fixes some boundary condition  $\az = A_B$ on its edge. 
The area of the cell must remain constant because the $\bu$ that advects $\az$ is incompressible (similarly, the average of $\az$ across the cell is fixed) --- we are  therefore interested in a solution of \cref{Eq:Helmholtz} that is as
compact as possible for a given $K_\perp$,  thereby providing the lowest energy (smallest $K_\perp$) for a given sized cell. This is afforded by a cylindrically symmetric cell, so we define 
$(r,\theta)$ as the polar coordinates centered on the cell, yielding 
$\az\propto J_{0}(K_\perp r)$ as the only $\theta$-independent solution that does not diverge as $r\rightarrow0$. 
Note that an arbitrary constant can be added to the solution by changing the gauge of $\az$, but this must be added directly into the original variational problem.

Collecting these constraints, we obtain the perfectly circular magnetic-vortex solution
\begin{equation}
\begin{cases}
   \az(r)=A_{0}J_{0}(K_\perp r),& r<r_c\\
   \az(r)= A_B, & r\ge r_c,
   \label{Eq:Monopole}
\end{cases}
\end{equation}
where $r_c$ is the radius of the cell, at which $A_{0}J_{0}(K_\perp r)=A_B$ (as required to satisfy the boundary condition). Note that the two constants $A_0$ and $K_\perp$ are determined through the fixed area of the cell, the initial wave-action anastrophy, and the boundary conditions (assuming $\az$ is continuous at the start of the relaxation this will not provide a third constraint). This leaves no freedom to
allow the first derivative of $\az$ (i.e., the magnetic field) to be continuous, leading 
to an inevitable $\bb$ discontinuity 
across the cell boundary and a strong ring of current surrounding the cell. These features are clearly observed in the simulation, as shown in Fig.~\ref{Fig:Alfven_Vortices}, where we zoom in on various observed cells and highlight the large boundary currents (right panel).

The solution \eqref{Eq:Monopole} corresponds to a particular case of so-called ``Alfv\'en vortex'' solutions \cite{Petviashvili92, Alexandrova08}, in particular the vortex ``monopole.'' As well as resulting from the
variational argument, they arise as nonlinear solutions of ideal incompressible MHD equations. 
Indeed, the Helmholtz equation \eqref{Eq:Helmholtz} can instead be obtained by assuming zero flow $\bm{u}_{\perp}=0$ and $k_z\ll k_\perp$, which gives, from the momentum equation \eqref{Eq:Equations},
\begin{equation}
    \lbrace  \az,\gradp^2 \az \rbrace = 0. \label{Eq:Bracket} 
\end{equation}
Any functional relation $\gradp^2 A_z = f(A_z)$ satisfies \eqref{Eq:Bracket}, which subsumes any solution of the Helmholtz equation \eqref{Eq:Helmholtz} and 
thus Eq.~\eqref{Eq:Monopole}. In this minimum energy, constant-anastrophy solution, the contours of $A_z$ and $\gradp^2 A_z$ are circularly symmetric with aligned  gradients, thus nullifying the Poisson bracket \eqref{Eq:Bracket} \cite{Daniel19}. This nonlinear solution involves the magnetic tension balancing the perpendicular pressure. 

The theoretical considerations presented above provide more than a qualitative explanation for the turbulent ``cellularisation'' observed. We fit the magnetic  eddies highlighted  in  \cref{Fig:Alfven_Vortices} using the functional form
\begin{equation}
   \az(r)=\tilde{A}_{z0}J_{0}(K_\perp r)\left(1-f(r)\right)+\tilde{A}_z(r_c),
\label{Eq:AV_fit}
\end{equation}
where $f(r)$ is the logistic function $f(r)=(1+e^{-\kappa(r-r_c)})^{-1},$
which is effectively a step function that accounts for finite diffusive effects through the  ``logistic steepness'' parameter $\kappa$. The result of such a fit is shown in  \cref{Fig:AV_Fit} and demonstrates that the structures observed are unequivocally the minimum-anastrophy Alfv\'en vortices \eqref{Eq:Monopole}.

\begin{figure}
\begin{center}
\includegraphics[width=1.0\columnwidth]{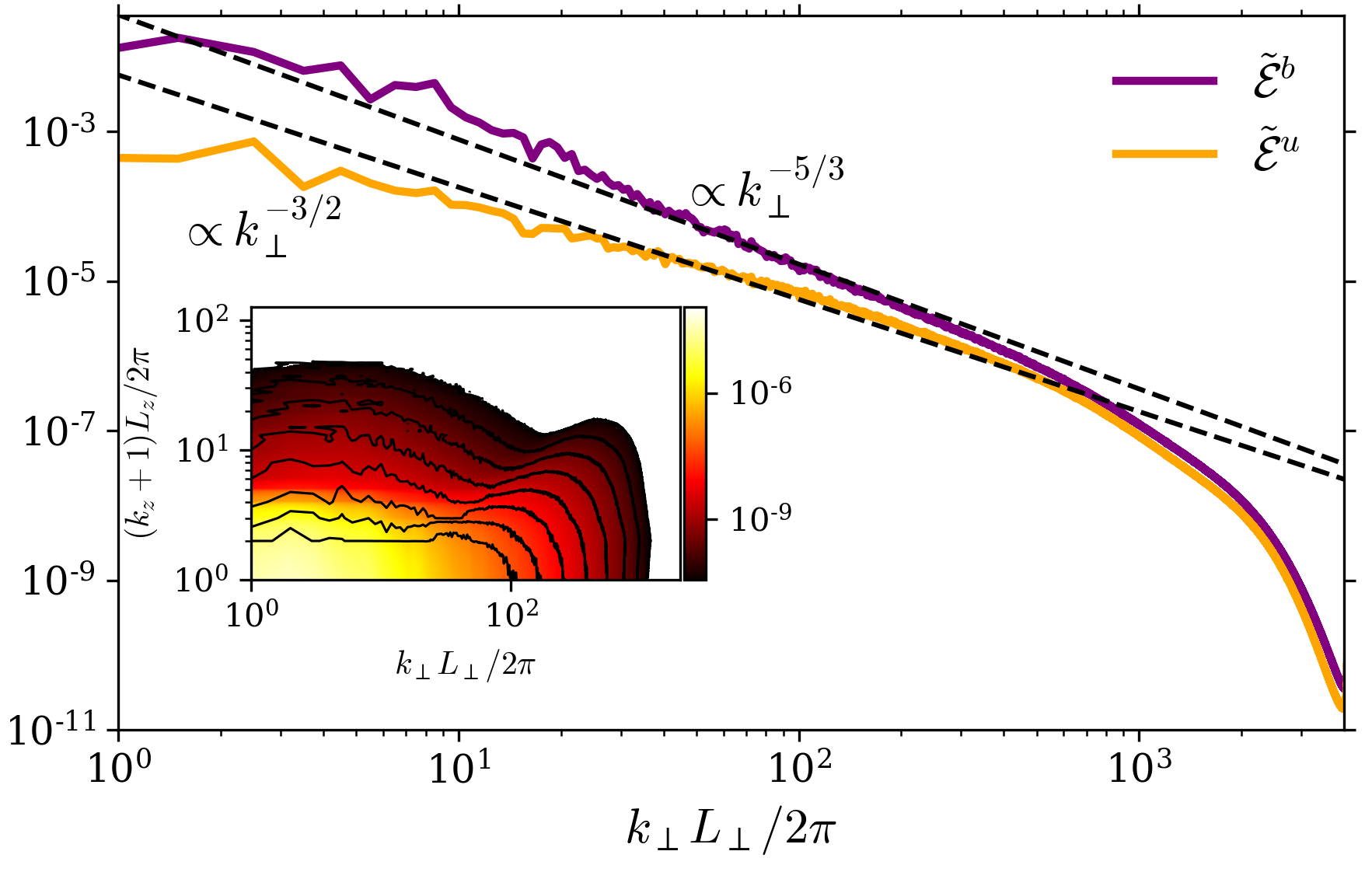}
\caption{Wave-action magnetic-energy spectrum $\tilde{\mathcal{E}}^{b}$ and kinetic-energy spectrum $\tilde{\mathcal{E}}^{u}$ at $a=250$ (cf. bottom panels of 
\cref{Fig:Snapshot}). The magnetic energy significantly dominates at large scales, with a steeper slope that eventually joins the velocity spectrum at small scales. The inset shows the 2D $k_{\perp},k_{z}$ spectrum of magnetic energy, illustrating how it is significantly dominated by the 2D $\Delta=0$ modes (the only expansion-dominated $\Delta<1/2$ modes in our domain).  }
\label{fig: balanced phase spectra}
\end{center}
\end{figure}

In \cref{fig: balanced phase spectra}, we illustrate the magnetic- and kinetic-energy spectra, $\tilde{\mathcal{E}}^{b}$ and $\tilde{\mathcal{E}}^{u}$ respectively. The strong magnetic dominance at large scales leads to a steeper magnetic spectrum, approximately $\tilde{\mathcal{E}}^{b}\propto k_{\perp}^{-5/3}$ at large scales, with a flatter velocity spectrum that eventually joins the magnetic spectrum at small scales. This is qualitatively similar to those observed at large scales during 
very low cross helicity periods in the solar wind \cite{Tu91}. The inset shows the 2D $k_{\perp},k_{z}$ magnetic-energy spectrum, illustrating the dominance of the 
$k_{z}=0$ 2D modes at these late times.  
The velocity fluctuations seem to be dominated by regions between the the individual  magnetic ``cells,''  arising from the coalescence of the Alfv\'en vortices through magnetic reconnection, which generate out-flows in the reconnection exhausts. 
The  Alfv\'en vortices thus slowly move around, thereby  generating further collisions. As the simulation progresses, ever larger magnetic structures are generated via mergers of  Alfv\'en vortices, creating further outflows that trigger more merging, thus minimizing the total energy and causing a slow nonlinear decay (this is overwhelmed by the expansion-induced growth). This hierarchical process, which is the basis of 2D MHD decaying turbulence theories \cite{Zhou19} is, however, impeded by the expansion, which acts as an additional damping of the outflows, 
hindering the nonlinear dissipation; indeed, the turbulent dissipation rate in our simulations, which is measured to scale as ${\sim}a^{-0.2}$ at late times, is slower than in 2D MHD \cite{Hatori1984,Zhou19}.
At large radial distances, these nonlinear effects therefore tend to ``freeze up''  and the structures become more and more static in time, growing at almost  the rate  predicted from linear theory (${\propto}a^{1}$).

The stability of such structures is an interesting question that we do not study in detail. The inevitability of the intense current rings suggests that at sufficiently high Lundquist number these ``ground state'' Alfv\'en vortices will become tearing unstable and break up into plasmoid chains confined on rotating rings. 
Allowing for the finite length of the structures and/or compressible effects may also lead to instabilities. Indeed, given the background mean field, the equilibrium \eqref{Eq:Monopole} is effectively a  screw pinch, with its nonlinear equilibrium resulting from the balance between the curvature/tension force of $\bb$ and the pressure gradient.
 Such equilibria can be unstable to kink instability and  sausage instabilities \cite{Kruskal54,Schuurman1969}. Thus, by assuming 
 the plasma to be incompressible with constant density, the RMHD model may 
miss important effects in their description, particularly instabilities that could aid in their destruction and
enhance nonlinear dissipation.

\section{Solar wind observations}\label{sec: relation to observations}

\subsection{Parameters and scales of the Solar wind}

Let us first remind the reader  that our simulation parameters
were chosen primarily to test and understand the dynamics  of 
reflection-driven turbulence, rather than simulate specific solar-wind streams. In particular,  the large $\chi_{\rm exp0}$  and small-scale $\zp$ (small $\Lp/L_\perp$) are  more extreme than occurs in the super-Alfv\'enic solar wind,  while the distance range (up to $a=1000$) is wider than regularly considered in observational studies. Via the phenomenology of \cref{sub: dmitruk decay}, this leads to a longer phase of imbalanced-turbulence decay  before the transition to the balanced phase, thus improving our analysis of these dynamics. Moreover, our highly idealized initial conditions are clearly inappropriate, since significant evolution will have occurred before reaching $R_{\rm A}$ in the sub-Alfv\'enic regions of the wind. 

We can estimate more realistic parameters using recent PSP results from $R\simeq R_{\rm A}$ \cite{Kasper2021}. Using $L_{+}\sim U/f_{\rm out}$, where $f_{\rm out}$ is a characteristic measured  frequency the energy-dominant scale (we take the spectral break in figure 2 of Ref.~\cite{Kasper2021}), and using the fact that $\va\simeq U$ at $R\simeq R_{\rm A}$, one
finds
\begin{align}
    \chi_{\exp} &= \frac{z^+_{\rm rms}/\Lp}{\dot{a}/a} \sim f_{\rm out} \frac{R}{U^2} {z^+_{\rm rms}} \nonumber\\ &\simeq 62 \frac{f_{\rm out}}{2\times 10^{-3}{\rm Hz}} \frac{R/(18 R_\odot)}{U/400{\rm kms}^{-1}}\frac{z^+_{\rm rms}}{\va}.\label{eq: solar wind chi exp}
\end{align}
This estimate ignores many uncertainties, including the difference between parallel and perpendicular scales,  differences between streams, and the violation of Taylor's hypothesis near the Alfv\'en point, but should 
at least give an order-of-magnitude estimate of the $\chi_{\rm exp}$ of the $z^+$ fluctuations that dominate the total energy.
We see that for observed fluctuation amplitudes, we expect $\chi_{\rm exp}\gg1$, but not nearly so large as the $\chi_{\rm exp0}$ chosen for our most extreme simulation. This further implies that
the transition into the balanced, magnetically dominated regime 
will occur at smaller radii than implied by \cref{Fig: time evolutions},
and the separations between the scales of $z^{+}$ and $z^{-}$, or between 
the initial and final scales of $z^{+}$, will be much smaller.

\subsection{Relation to specific observations in the solar wind}

Here we outline various predictions of  reflection-driven turbulence that can be directly 
compared to solar-wind observations. We particularly focus on the importance of $\chi_{\rm exp}$, including
the possibility that a correlation of $\chi_{\rm exp}$ with wind speed could naturally 
explain numerous other well-known observational correlations.


\subsubsection{Imbalance evolution}\label{subsub: imbalance observations}

There has been substantial previous literature devoted to understanding
the observed evolution of imbalance (normalized cross helicity) with radius, as well as its correlation with wind speed \cite{Roberts1987,tumarsch95}.
A particular focus has been understanding why the imbalance is observed to decrease with increasing radius in the solar wind, even though decaying MHD turbulence simulations (and theory) robustly show (and predict) the opposite \cite{Dobrowolny1980a,Grappin1982,Chen2011a,Schekochihin2022}. Some suggestions 
invoke interactions between different streams as the dominant influence
\cite{Roberts1992,Matthaeus2004,Adhikari2015}, or parametric decay of Alfv\'en waves \cite{Goldstein1978}, but we see that the imbalance decrease is naturally explained by reflection without invoking any other physics. 
While we are certainly not the first to suggest this \cite{Velli89,velli93,Hollweg2007,Verdini09,Grappin2022}, our simulations and phenomenology 
clarify why this occurs  and provide simple, testable predictions that (to the best of our knowledge) have not appeared in previous literature. 

As argued above, the key parameter governing the imbalanced decay phase is $\chi_{\rm exp}$, the ratio of nonlinear to expansion rates. 
The basic phenomenology of \cref{sub: dmitruk decay} \cite{Dmitruk02,verdini07} predicts $\zm \sim \zp/\chi_{\rm exp}$, with 
$\chi_{\rm exp}\sim (z^{+}/\lambda_+)/(\dot{a}/a)$ seen to scale as $\chi_{\rm exp}\propto a^{-3/2}$ in 
our simulations where $\tilde{E}^{+}\sim a^{-1}$ and $\lamp\sim a^{1/2}$ (as inferred from the evolution of $\zm$). This implies that $\sigma_{c}$ evolves as 
\begin{equation}
\sigma_{c} \sim \begin{cases}
\frac{1-\chi_{\rm exp}^{-2}}{1+\chi_{\rm exp}^{-2}}, & \chi_{\rm exp}\gtrsim 1\\
0 , & \chi_{\rm exp},\lesssim 1
\end{cases},
\end{equation}
which, for  $\chi_{\rm exp} \sim \chi_{\rm exp0}a^{-3/2}$, stays  in a highly imbalanced state near 
$\sigma_{c}= 1$ across a wide range of $a$ before rapidly dropping towards zero around the radius 
$a\sim \chi_{\rm exp0}^{2/3} $ (the exact power-law exponent $-3/2$ makes no difference to  this basic picture). The model 
thus naturally explains the observed  radial 
dependence of the turbulence from imbalanced to balanced.  Similarly, the observed differences between fast and slow streams would be well explained if fast-wind streams start with  larger $\chi_{\rm exp0}$ around $R_{\rm A}$ (and therefore also maintain larger $\chi_{\rm exp}$ throughout their evolution). This prediction is  easily testable observationally. It is  also physically expected based on reflection-driven 
models of the sub-Alfv\'enic regions \cite{Wang1990,cranmer07,Cranmer2009,Chandran2021}, in which 
slower streams arise because more of the outward-fluctuation energy
is dissipated at low altitudes \cite{Halekas2023}, thus giving a lower amplitude at large radii and therefore a lower $\chi_{\rm exp}$.
For the stream  with $\chi_{\rm exp0}\simeq60$  discussed
above (\cref{eq: solar wind chi exp} \cite{Kasper2021}),  evolution with $\chi_{\rm exp}\propto a^{-3/2}$ predicts that the fluctuations should reach small $\sigma_{c}$ around $a\approx 15$,
or a little beyond $1{\rm AU}$ --- certainly not unreasonable.

\subsubsection{The $\sigma_c$-$\sigma_r$ ``circle plot''}

The radial  evolution of $\sigma_{c}$ and $\sigma_{r}$ on the ``circle plot'' of \cref{Fig: time evolutions} provides simple, persuasive 
evidence that our reflection-turbulence model captures key aspects of solar-wind evolution. 
Observations robustly show that solar-wind  fluctuations are concentrated near the circle's bottom-right quadrant edge, evolving from 
$(\sigma_c,\sigma_r) =(1,0)$ close to the sun towards  $(\sigma_c,\sigma_r) =(0,-1)$ at large radii \cite{Bavassano1998,Bruno07,Bruno2013,Wicks13}.
This behavior agrees with 
our simulations and phenomenological arguments (\cref{Fig: time evolutions}): during its imbalanced phase, the system 
remains close to the circle's edge because reflection  generates $\bm{z}^{-}$ fluctuations that are anti-aligned with 
their $\bm{z}^{+}$ source (negative $\sigma_{\theta}$; see \cref{Eq:Cos}), evolving into the $\sigma_{r}\simeq-1$ state at late 
times as the long-wavelength expansion-dominated modes start dominating the dynamics (\cref{sec: Balanced phase}).
In addition,  faster wind is observed to be concentrated near the middle right (large $\sigma_{c}$, small $\sigma_{r}$), while  
slower streams are concentrated near the bottom (large $\sigma_{r}$, small $\sigma_{c}$) \cite{Bavassano1998,Bruno07}, which fits straightforwardly into the idea described above that fast-wind streams have larger $\chi_{\rm exp}$, thus spending longer in the imbalanced phase. 
While we are, again, not the first to speculate on the relevance of expansion to these observations \cite{Hollweg2007,Grappin2022}, we believe ours are the first simulations to highlight this feature, particularly the evolution into the balanced
$\sigma_{r}\simeq-1$ state and the importance of the Els\"asser alignment $\sigma_\theta$.

\subsubsection{Fluctuation spectra and the $1/f$ range}

Many years of observations have shown that magnetic 
fluctuations in the solar wind display a $1/f$ slope at low frequencies, 
differing from the steeper $f^{-3/2}$ or $f^{-5/3}$ scalings 
observed at higher frequencies in the inertial range \cite{Goldstein1984,Matthaeus86,Chen20}. There is currently no consensus on the origin of this $1/f$ range. Suggestions range  from its origin in the low corona \cite{Matthaeus86}, implying that it is the energy reservoir that feeds the solar-wind turbulent cascade \cite{Matthaeus94}, to it being 
the  result of  spherically polarized fluctuations growing to amplitudes larger than one \cite{Matteini18}, or parametric decay of compressive fluctuations \cite{Chandran18,Huang2023,Davis2023}. Numerous studies \cite{Velli89,Verdini08,Verdini09}
have also shown  that reflection-driven 
turbulence can naturally create $1/f$ spectra in both the parallel \cite{Verdini12} and perpendicular \cite{Perez13,Chandran19} directions. Our results 
agree with the latter\footnote{Since the  RMHD model is unsuitable for capturing large-amplitude fluctuations the parallel spectra at these large scales should not be believed.} through the mechanism described in Refs.~\cite{Velli89,Perez13} (see \cref{sub: imbalanced spectra}). In line with previous works \cite{Perez13,Chandran19}, 
we find that the spectral scaling of  $\zm$ is steeper than that of
$\zp$ through this range, scaling as $\tilde{E}^{-}\propto k_{\perp}^{-3/2}$ in our simulations; this is similar (though
not identical) to  that observed \textit{in situ} \cite{Tu1990,Wicks13}, although this general signature is not
 unique to the reflection-turbulence model  \cite{Matteini18,Chandran18}. 
In addition, since the $1/f$ range in the model is generated by the turbulence during the imbalanced 
phase, at similar radii, we would expect $\chi_{\rm exp}\gtrsim 1$ regions  to exhibit a wider $1/f$ range than
$\chi_{\rm exp}\lesssim  1$ regions. If we further apply the hypothesis discussed above, that fast-wind streams have higher $\chi_{\rm exp0}$  than slow-wind streams, this would naturally explain  the well-known observation that the size of the  $1/f$ range correlates with wind speed \cite{Tu1989,Tu1990}.
In this context, 
it is also worth clarifying that the extremely wide $k_{\perp}^{-1}$
range seen in \cref{fig: imbalanced spectra} is again 
a consequence of the extreme parameters of the simulation (its
small initial scale and long imbalanced phase).
Finally, the general ideas naturally explain the results of Ref.~\cite{Wicks13} that in those (rare) regions with 
$\sigma_r\simeq\sigma_c\sim 0$, there is no significant $1/f$ range, since (given \cref{Fig: time evolutions}) such regions are presumably strongly influenced by physics that is unrelated to reflection-driven turbulence. 

\subsubsection{Inverse energy transfer and the split cascade}

The most significant qualitative difference between our energy 
spectra and previous results is the inverse energy transfer of $\tilde{E}^{+}$ caused by anomalous growth of wave-action anastrophy. 
This forces the decay to proceed via a split cascade, shifting the  $\tilde{E}^{+}\propto k_{\perp}^{-1}$ range to larger 
scales with time in the co-moving frame as it grows out of  a positive-slope infrared spectrum  at yet larger scales.
The feature is interesting in light of recent observations showing that the $1/f$ spectrum does not extend to the largest available scales,
especially near the sun \cite{Kasper2021,Huang2023}, instead developing as the wind propagates outwards 
\cite{Davis2023}. The surprising, non-trivial prediction of our model
is that the correlation scales of the fluctuations, which 
lie towards large-scale side of the $1/f$ range, should increase with $R$ faster than expansion (i.e., increase in the co-moving frame). In addition, the split cascade itself may be directly observable, with the 
prediction that the cascade of $z^{+}$ should switch from forward to inverse at the largest scales in imbalanced turbulence (see \cref{Fig:Fluxes}). Interestingly,  back transfer of energy from small to large scales has been observed in $z^{+}$ in highly imbalanced streams \cite{Smith09,Coburn15},
although since these observations seem to pertain to smaller scales
(where we observe a forward cascade of both $z^{+}$ and $z^{-}$; \cref{Fig:Fluxes}), they may be unrelated.

This inverse energy transfer  could have broader implications for  solar-wind turbulence and acceleration, particularly if similar physics also applies in sub-Alfv\'enic regions. 
Close to the Sun, the  large gradient of the Alfv\'en speed around the transition region should prevent low-frequency 
Alfv\'en waves launched from the chromosphere from propagating outwards to large radial distances \cite{Hollweg1978,Leroy1981,Velli91,Reville18}. If the chromospheric fluctuations are turbulent and critically balanced, with little power in modes with $\va k_\|>z^{\pm} k_\perp$ \cite{Schekochihin2022}, this 
high-pass filter  would  also have the effect of filtering out large scales in the perpendicular direction, leading to small correlation 
scales at the coronal base. 
The fact that low-frequency waves end up 
dominating the solar wind spectrum is therefore highly non-trivial and naturally suggests that some form of  inverse energy transfer 
is needed to explain the existence of large-scale fluctuations at all. 
The anomalous growth of wave-action anastrophy could provide one such mechanism.

\subsubsection{Solar wind heating}

Our study also has  application to the understanding of solar-wind heating, 
although more work is needed. In fast-wind streams, the observed radial decrease of the proton 
temperature $T$ is slower than 
in adiabatic cooling, indicating that 
the plasma is heated as it moves outward from the sun \cite{Marsh82,Freeman88,tu88}, presumably 
by the dissipation of fluctuations \cite{Verma1995,Vasquez2007,Chen20,Halekas2023}. 
Such turbulent heating appears to be less important in slower-wind streams, although results remain controversial \cite{Freeman88,Totten1995,Hellinger2011}. 
Within the RMHD EBM model, the heating rate per unit volume is $Q = -\rho \langle \bm{z}^{+}\cdot\bm{D}^{+}+\bm{z}^{-}\cdot\bm{D}^{-}\rangle/2$ ,
where $\bm{D}^{\pm}$ represents the hyper-viscous terms included to dissipate energy at small scales (\cref{Eq:Hyper-visc}). 
Converting to wave-action variables  and using the total energy conservation \eqref{Eq: wave action energy def} (see also Ref.~\cite{Perez2021}) gives
\begin{equation}
Q = - \rho\frac{\dot{a}}{a}\left( \frac{\partial \tilde{E}}{\partial a} + \frac{\tilde{E}^{r}}{a}\right).\label{eq: heating rate}
\end{equation}
During the imbalanced phase at high $\chi_{\rm exp}$, when $\tilde{E}^{r}\ll \tilde{E}\approx \tilde{E}^{+}$, the phenomenology of \cref{sub: dmitruk decay} predicts  $\tilde{E}^{+}\propto a^{-1}$ and $\rho\propto a^{-2}$, so that $Q\propto a^{-5}$. 
Then, as the system transitions into the balanced phase, the heating rate drops significantly as the system becomes dominated by slowly evolving Alfv\'en vortices. 
At late times we measure a small residual nonlinear  dissipation that causes $\tilde{E}\propto a^{0.8}$ (rather than the  $\tilde{E}\propto a^{1}$ predicted
by linear theory), implying a heating rate that flattens to $Q\propto a^{-3.2}$. 

Whether these results are consistent with observations remains
unclear. Most observational studies have inferred heating rates by fitting power-law profiles to the observed temperatures, then comparing the inferred scalings to ``adiabatic'' profiles that would occur in the absence of heating: $T\propto R^{-4/3}$ for an isotropic fluid (i.e., if the perpendicular and parallel temperatures are well  coupled, $T_\perp\sim T_\|$), or $T_\perp\propto R^{-2}$ for a collisionless plasma (or more generally, $T_\perp\propto B$). 
A $Q\propto a^{-\alpha}$ heating 
profile with $\alpha\geq 5$ ($\alpha\geq 13/3$ for an isotropic fluid) is too steep to lead to a power-law temperature profile that differs from the adiabatic profile at asymptotically large $R$; however, depending on the magnitude of $Q$,
almost any local scaling of $T$ can be realized (it just does not vary as a power law over a wide range in $R$). This, combined with the effects of averaging over different streams with different $\chi_{\rm exp}$, makes it unclear whether the difference between 
the high-$\chi_{\rm exp}$ prediction ($Q\propto a^{-5}$)  and the classic result that $\alpha \approx 3.8\rightarrow 4$ \cite{Freeman88,Totten1995,Hellinger2011} should signal the importance of other physics, or not. 
Indeed, more complex models based on a similar phenomenology \cite{Cranmer2009,Chandran2011}, 
reproduce observed temperature profiles reasonably well out to ${\simeq}1{\rm AU}$.
Also of interest is the transition around $\chi_{\rm exp}\sim 1$, where we predict that the 
decrease in heating rate with radius should slow to eventually approach $Q\propto a^{-3}$ as 
the heating stops. If slow-wind streams have smaller $\chi_{\rm exp}$ as suggested above, 
the general trend could be consistent with the observation of closer-to-adiabatic evolution 
in slow wind (a flatter power-law profile of $Q$ will not be measurable if its magnitude is too small), as well as recent measurements showing the much greater importance of wave heating in fast, compared to slow streams \cite{Halekas2023}.

Overall, while plausibly consistent, more work is needed, particularly to quantify the
relevance reflection-driven turbulence compared to other effects including  pick-up ions at larger radii \cite{Gazis1994} and stream interactions \cite{Roberts1992} or parametric decay \cite{Goldstein1978,Chandran2018,Shoda2019} in highly imbalanced regions.

\subsubsection{Alfv\'en vortices}

The final phases of our simulations are characterized by isolated magnetically dominated nonlinear solutions (Alfv\'en vortices), 
in which magnetic tension balances the total pressure. These structures are dominated by expansion, so not
expected to be critically balanced in the usual sense (in our simulations they are truly 2-D), with  sharp boundaries and current rings that separate them from the surrounding more quiescent plasma (see \cref{Fig:Alfven_Vortices}). 
Although the Parker spiral and  compressible and finite-amplitude effects  are likely  important for their evolution, 
we suggest that they provide a natural explanation for  the so-called 
``Magnetic Field Directional Turnings'' (MFDTs) observed in the solar wind at large radii, whose origin has challenged a clear theoretical explanation so far  
 \cite{Tu91,Bruno07}. The observed structures are highly magnetically dominated, with an approximate balance between thermal and magnetic pressure and very sharp boundaries in $\bm{B}$  \cite{Tu91}, just as observed in \cref{Fig:Alfven_Vortices}.    This explanation suggests that  MFDTs and Alfv\'en vortices \cite{Petviashvili16, Alexandrova08} are the same physical entity whose origin is reflection-driven turbulence. It could be tested by several means such as: (i) a direct fit observed structures at large radii to \cref{Eq:Monopole}, perhaps with a focus on high-latitude regions where the Parker 
spiral is less dominant; (ii) by examining their parallel scales, which should satisfy $\Delta\lesssim 1/2$, thus indicating they are expansion dominated; and  (iii) 
by examining the radial dependence of $\delta B_{\perp}/\overline{B}$, which should grow ${\propto}R$ until reaching large amplitudes ($\delta B_{\perp}/\overline{B}\sim1$, where our RMHD approximation is no longer valid).

\section{Conclusion}\label{sec: conclusion}

This work has presented a detailed computational and phenomenological 
study of  reflection-driven turbulence, which is thought to play a key role in the heating and acceleration of the solar wind \cite{Cranmer2009}, as well
as in other magnetized, highly stratified environments such as accretion disk coronae \cite{Chandran2018}.
We have approached the problem from the simplest standpoint possible, using the reduced-MHD expanding box model (EBM), which
captures Alfv\'enic (incompressible and perpendicular) dynamics and assumes  a constant wind speed $U$ that is faster than the Alfv\'en speed $\va$.
By enabling very high simulation resolutions and clarifying the analysis, this 
has helped to reveal a rich and nontrivial dynamics that displays
features reminiscent of both forced- and decaying-turbulence paradigms.
In order to explore these features in depth, our study has differed from most previous works 
by deliberately not attempting to match solar-wind parameters, instead focusing 
on understanding the basic physical processes. 
While highly idealized, our results can plausibly explain a  range  of disparate observations from \textit{in-situ} spacecraft (see \cref{sec: relation to observations}), 
giving us some confidence in the value of the computational approach and the utility of the theoretical framework.

Our most surprising novel results relate to the existence of strong inverse energy transfer, with 
the decay of the dominant  outward-propagating fluctuations proceeding via a split cascade that transfers energy to small 
and large scales simultaneously. We argue that this results from an anomalous conservation law of the ``wave-action anastrophy'' $\anas$
(the box-averaged parallel vector potential squared), which can grow due to the effects of expansion in the strongly turbulent 
system. We provide a heuristic theoretical argument justifying this (\cref{sub: anastrophy})
based on linear-wave dynamics and the observation that the Els\"asser fields $\bm{z}^{\pm}$ remain nearly aligned ($\bm{z}^{-}\propto -\bm{z}^{+}$) and anomalously coherent (effectively propagating in the same direction). This latter property, which results from the suppression of collisions between Alfv\'enic fluctuations, as diagnosed in the simulation via frequency spectra (\cref{sub: anomalous coherence}), 
leads to a turbulent decay that remains strong even 
though the minority fluctuations ($\bm{z}^{-}$) have very low amplitude \cite{Verdini09,Perez13}. 
Using these core ideas, the radial evolution of the energy, imbalance (normalized cross helicity), and residual energy
are analysed via a heuristic phenomenology based on previous works \cite{Dmitruk02,verdini07,Chandran09a}, 
 extended to account for the radial evolution of the different scales of $\bm{z}^{\pm}$ (\cref{sub: dmitruk decay}). 
We argue that a key  parameter is $\chi_{\rm exp}$, which, as the ratio of the expansion/reflection timescale  to the nonlinear timescale,
 naturally controls the reflection-driven turbulent decay. 
Overall the phenomenology provides a reasonable match to most simulation results, although there remain some unresolved discrepancies. 

A secondary result of our work concerns the long-term evolution of the system at large radii, as relevant 
to the outer heliosphere and regions of slower wind (see below). Our simulations show that  super-Alfv\'enic reflection-driven 
turbulence is characterized by two distinct phases, separated by the radius at which $\chi_{\rm exp}\sim1$ where
the system becomes balanced ($z^{-}\sim z^{+}$) and dominated by long-parallel wavelength modes 
for which expansion overwhelms the Alfv\'enic restoring force. From this radius onwards, nonlinear 
interactions slow significantly as the system cellularizes  into a collection of nonlinear ``Alfv\'en vortex'' solutions separated by 
sharp current-ring boundaries. The structures, which are strongly magnetically dominated, slowly move and merge while their normalized  amplitude $|\bm{B}_\perp|/|\overline{\bm{B}}|$ grows rapidly due to 
expansion. 

\subsection{Observations}

Despite the simplicity of our RMHD expanding box  and the many important physical effects that are unjustifiably neglected (see \cref{sub: uncertainties conclusion} below), 
its predictions seem to explain a range of different well-known solar-wind observations. In \cref{sec: relation to observations}, 
we outline a number of these ideas in a way that should be understandable without a detailed reading of the main text, 
as well as making more specific predictions that may help to further test and refine the reflection-driven turbulence paradigm. 
In summary, the model naturally explains the observed decrease in turbulence imbalance with heliocentric radius \cite{tumarsch95}, as 
well as its correlation with wind speed if $\chi_{\rm exp}$ is statistically lower in slower streams, as expected from flux-tube expansion models \cite{Chandran2021}.  
For similar 
reasons, observations of the classic $\sigma_{c}$-$\sigma_{r}$ ``circle plot''   \cite{Bavassano1998} are reproduced numerically 
and understandable by appeal to the simple phenomenology and the dominance of long-wavelength structures in the balanced regime.
The transition from imbalanced to balanced turbulence also entails a slow shutting off of the turbulent heating, 
which seems plausibly consistent with observations of radial and stream dependence of solar-wind heating rates \cite{Totten1995,Halekas2023}, though more detailed models and observations are needed \cite{Chandran2011}. 
Our simulations, as well as previous literature on the subject \cite{Velli89,Verdini09,Perez13}, reproduce the well-known $1/f$-range spectrum at large scales ($\mathcal{E}^{+}(k_{\perp})\propto k_{\perp}^{-1}$ in the simulation). Because of the inverse energy transfer, this forms naturally from smaller-scale fluctuations 
in the initial conditions, migrating to larger scales in the co-moving frame with time. This inverse energy transfer may be observable through its 
radial dependence or via direct measurements of the turbulent flux, and could 
have interesting consequences for explaining the dominance of low-frequency fluctuations in observations, 
even though they should be filtered out by the
large Alfv\'en-speed gradients in the upper chromosphere.
Finally, the  magnetically dominated Alfv\'en vortices, which inevitably dominate the solutions at large $a$,  seem to bear close resemblance to ``Magnetic Field Directional Turnings'' \cite{Tu91},
which are observed at large heliocentric distances.

\subsection{Uncertainties and future work}\label{sub: uncertainties conclusion}

Due to both the highly idealized model and the details of the simulation design, our study 
is beset with a number of significant  uncertainties. While we do not believe that these fundamentally 
invalidate our 
main results, they are nonetheless important to acknowledge and, hopefully, to rectify in future work. 

The basic phenomenology of \cref{sub: dmitruk decay} \cite{Dmitruk02} does not satisfactorily 
explain some features of the the imbalanced-phase turbulence, and a priority of future work should be 
to understand this ``platonic'' form of reflection-driven turbulence in the expanding box. Of particular difficulty 
is the relationship between the growth  of $\zm$, which we observe to be significantly (${\propto}a^{1/2}$) faster  than the standard 
prediction \cite{verdini07,Chandran09a}, and the evolution of the dominant scales of $\zp$ and $\zm$ ($\lamp$ and $\lamm$). The growth of $\lamp$ and faster-than-expected 
growth of $\zm$  accelerate the transition into the balanced regime, thus decreasing the overall energy decay and heating, 
so these uncertainties pertain directly to the global energetics of the solar wind. 
It may be that some of these discrepancies with the model relate to our initial conditions, and indeed
we have found some dependence of the results on the initial conditions (e.g., the infrared spectrum and  parallel scales)  that 
remain incompletely understood.  Another 
important goal of future work should be to better explore the dependence on $\Delta_{\rm box}$, which 
sets the range of parallel wavelengths available to the system. In our simulations, which fixed $\Delta_{\rm box}=10$, 
only the $k_{z}=0$ 2-D mode is linearly expansion dominated (non-Alfv\'enic), but in reality there 
should be a continuum of such modes down to the scales where global effects become important ($k_{z}\sim 1/R$).
Decreasing $\Delta_{\rm box}$ is equivalent to  increasing the parallel box scale $L_{z}$, and therefore increases the 
simulation cost, but this should be explored in future work. An additional priority for future work is to elucidate the physical mechanisms that give rise to the
$E^+(k_\perp) \propto k_\perp^{-1}$ and $E^-(k_\perp)\propto k_\perp^{-3/2}$
scalings shown in the bottom panel of figure~\ref{fig: imbalanced spectra}, 
 which are not explained by existing cascade models for imbalanced MHD turbulence \citep[e.g.][]{Velli89,Lithwick07,Chandran19}.

Moving beyond the uncertainties in interpreting the RMHD EBM results, there exist many uncertainties related to the 
model itself. Although its simplicity is appealing, RMHD obviously cannot capture any compressive physics or the physics
of the large-amplitude spherically polarized fluctuations that are routinely observed \textit{in situ} \cite{Bale2019}. The 
latter can be rectified via full MHD simulations \cite{Squire2020}, but the former arguably cannot, given that the solar wind is 
a collisionless plasma with compressive fluctuations that may or may not be well described by fluid models \cite{Schekochihin2009,Verscharen2017}.
These issues, as well as our neglect of the Parker spiral, are likely particularly important for our results related to 
Alfv\'en vortices, since these structures are inherently compressive (though in total pressure balance).
There also exist various subtle issues related to the EBM, motivating future studies with global flux-tube models \cite{vanBallegooijen2011,Perez13} that are more focused on super-Alfv\'enic regions. The EBM should accurately capture dynamics 
only in the limit where a reflected $\bm{z}^{-}$ cannot propagate further than one box length, because otherwise 
this $\bm{z}^{-}$ could re-encounter the same $\bm{z}^{+}$ multiple times (clearly an unphysical effect). This likely 
limits its applicability to study of the strong-turbulence regime where $\bm{z}^{-}$ is anomalously coherent. Another effect that
cannot be captured in the EBM due to its fixed parallel size relates to the increased range of 
long-wavelength, expansion-dominated modes that become available to the system at increasing radius as it transitions into the 
balanced regime  (see  \cref{sub: linear balanced}).

Finally, a key omission, which has been made purely for the sake of simplicity, is the recently 
discovered ``helicity barrier'' effect \cite{Meyrand21}. The helicity barrier suppresses dissipation 
via electron heating due to finite-Larmor-radius effects in $\beta\lesssim 1$ turbulence, channeling the 
turbulent flux into ion-cyclotron heating only once the fluctuations can reach sufficiently small parallel scale \cite{Squire2022}.
By suppressing the dissipation of $z^{+}$, the helicity barrier could significantly change our results 
in $\beta\lesssim 1$ regions, bringing in direct dependence on the parallel scales. Therefore our
results here can only apply  to either the saturated phase, in which the energy flux into ion-gyroradius scales 
is balanced by ion heating through the cyclotron resonance \cite{Bowen2023}, or to $\beta\gtrsim 1$ regions.
 Understanding the impact of the helicity barrier on reflection-driven turbulence should be a priority for future work. 
 
\acknowledgements
The authors would like to acknowledge enlightening conversations with Roland Grappin, Andrea Verdini, Chris Chen, and Alex Schekochihin.
RM and JS acknowledge the support of the Royal Society Te Ap\=arangi, through Marsden-Fund grants MFP U0020 (RM) and MFP-UOO2221 (JS), as well as through the Rutherford Discovery Fellowship  RDF-U001804  (JS).
BC was supported in part by
NASA grant NNN06AA01C to the Parker Solar Probe FIELDS Experiment and by NASA grant 80NSSC19K0829. We wish to acknowledge the use of New Zealand eScience Infrastructure (NeSI) high-performance computing facilities as part of this research. New Zealand's national facilities are provided by NeSI and funded jointly by NeSI's collaborator institutions and through the Ministry of Business, Innovation \& Employment's Research Infrastructure programme. We also wish to acknowledge the generous hospitality of the Wolfgang Pauli Institute, Vienna, where these ideas were discussed during several `Plasma Kinetics' workshops.  

\label{APPENDIX STARTS HERE}

\appendix

\section{Anastrophy dissipation in linear waves}\label{app: linear anastrophy}

In \cref{sub: anastrophy}, we argued that anomalous wave-action anastrophy growth 
places a strong constraint on reflection-driven turbulent dynamics, forcing the fluctuations
to rush towards larger scales as they decay. As part of this argument, we pointed 
out that linear propagating waves with $\Delta >1/2$ ($k_{z}\vao>\dot{a}/2$) are particularly 
efficient at destroying anastrophy via the term $\langle \zetap \partial_z \zetam\rangle$. The 
 corollary is that a system with either (i) smaller $\zm/\zp$ than a linear wave, or (ii) wave phases that 
 are scrambled compared to the linear wave, will grow wave-action anastrophy faster  than the 
 linear (dissipationless) system. In this appendix, we  examine the cause
 of this linear anastrophy dissipation by computing $\langle \zetap \partial_z \zetam\rangle$ for a generic collection of linear waves, demonstrating explicitly how it cancels the wave-action anastrophy growth term ($a \anas$ in Eq.~\eqref{Eq: anastrophy conservation}). Of course, this is no surprise --- given that
 $\anas$ does not grow on average in a linear propagating ($\Delta>1/2$) wave, it is inevitable --- nonetheless, 
 aspects of the calculation are interesting and  worth presenting. 

 The potentials $\zetapm$ evolve according to effectively  the same linear equation 
 as $\zpm$ (see \cref{sub: linear balanced}):
 \begin{equation}
\dot{a}\frac{\partial\zetapm}{\partial \ln a}   =  \pm {\vao}\frac{\partial \zetapm}{\partial z} - \frac{\dot{a}}{2} \tilde{\zeta}^{\mp}.\label{Eq: elasser potentials linear}
\end{equation}
Assuming plane waves with $\ln a$ as the time variable, $\zetapm (\bm{x},t) = \zetapm_{\bm{k}} e^{i \bm{k}\cdot \bm{x} - i \omega \ln a}$, 
the  linear eigenfrequencies  are \cref{Eq:Eigenfreqs} ($\omega^\pm = \pm \sqrt{\Delta^2-1/4}$) with eigenmodes $\xi^\pm_{\bm{k}} = \zetapm_{\bm{k}}/2 \pm i \zetamp_{\bm{k}} (\Delta - \sqrt{\Delta^2-1/4} )$. 
Inverted, this latter expression gives
\begin{equation}
    \zetapm_{\bm{k}} = 2\frac{\xi^\pm_{\bm{k}} \mp 2 i \Theta \xi^\mp_{\bm{k}}}{1-4\Theta^2} = f^\pm_{\bm{k}} \xi^+_{\bm{k}} + g^\pm_{\bm{k}} \xi^-_{\bm{k}},\label{Eq:app: zeta from emodes}
\end{equation}
where $\Theta \equiv \Delta - \sqrt{\Delta^2-1/4}< 1/2$ for $\Delta >1/2$,  with $\Theta \approx (8\Delta)^{-1}$ for $\Delta \gg1$, and the $f^\pm_{\bm{k}}$ and $g^\pm_{\bm{k}}$ coefficients are defined for notational convenience below.
Taking general  initial conditions $\zetapm_{0,\bm{k}}$ (equivalently  $\xi^\pm_{0,\bm{k}}$), we compute the right-hand side of the  anastrophy equation \eqref{Eq: anastrophy conservation}, to 
give
\begin{align}
\frac{\vao}{2}&\langle \zetap \partial_z \zetam\rangle = -\frac{\vao}{2}\sum_{\bm{k}}i k_z\left(f^+\xi^+_{0,\bm{k}} e^{i\omega^+t}+g^+\xi^-_{0,\bm{k}} e^{i\omega^-t}\right) \nonumber \\ &\qquad\qquad\qquad\qquad \times\left(f^-\xi^+_{0,\bm{k}} e^{i\omega^+t}+g^-\xi^-_{0,\bm{k}} e^{i\omega^-t}\right)^*\nonumber  \\
=& -\frac{\vao}{2}\sum_{\bm{k}} ik_z \left[f^+_{\bm{k}}(f^-_{\bm{k}})^* |\xi^+_{0,\bm{k}}|^2 +   g^+_{\bm{k}}(g^-_{\bm{k}})^* |\xi^-_{0,\bm{k}}|^2\right]\nonumber \\
=& \vao\!\!\! \sum_{\bm{k}_\perp,k_z>0} k_z\,{\rm Im}\left[ f^+_{\bm{k}}(f^-_{\bm{k}})^* |\xi^+_{0,\bm{k}}|^2 +    g^+_{\bm{k}}(g^-_{\bm{k}})^* |\xi^-_{0,\bm{k}}|^2\right]. \label{Eq: anastrophy damping linear} \end{align}
To arrive at the second line, we have additionally averaged over (or ignored) the wave periods to  eliminate
the rapidly oscillating  cross terms (${\propto}e^{2i\omega^\pm}$), which will cause the anastrophy to oscillate but cannot affect its longer-term evolution. Physically, this shows that 
any linear evolution necessarily picks up a correlation between $\zetap$ and $\partial_z\zetam$ (proportional to ${\rm Im}[ f^+_{\bm{k}}(f^-_{\bm{k}})^*]$ and ${\rm Im}[ g^+_{\bm{k}}(g^-_{\bm{k}})^*]$) because the eigenmodes $\xi^\pm$, which 
propagate in the $\pm \zh$ direction, contain both $\zetap$ and $\zetam$.
From Eq.~\eqref{Eq:app: zeta from emodes}, we see that $f^+_{\bm{k}}(f^-_{\bm{k}})^* = g^+_{\bm{k}}(g^-_{\bm{k}})^* = - 8i \Theta/(1-4\Theta^2)^2$, 
which (being imaginary and negative) shows that this correlation 
is such that linear waves are \textit{maximally efficient} at destroying  anastrophy (for a 
given magnitude of $\zetapm$). The obvious corollary is 
that if the phase of $\zetam$  is modified compared to that of $\zetap$ by  reflection-driven turbulence
(or anything else), the wave-action anastrophy will be destroyed less efficiently than it is
in a linear wave (again, for a given magnitude of $\zetapm$).

One can continue the calculation to work out the magnitude of \eqref{Eq: anastrophy damping linear}, but this calculation is most illuminating if we focus on the specific case of $\Delta\gg1$ and $\zetam_{0,\bm{k}}=0$. These imply $\xi^+_{0,\bm{k}} = \zetap_{0,\bm{k}}/2$, $\xi^-_{0,\bm{k}} = -i\Theta \zetap_{0,\bm{k}}\approx -i \zetap_{0,\bm{k}}/8\Delta$, such that  $|\xi^-_{0,\bm{k}}|^2\ll|\xi^+_{0,\bm{k}}|^2$ can be ignored in  \eqref{Eq: anastrophy damping linear}. 
Thus, 
\begin{align}
    \frac{\vao}{2}&\langle \zetap \partial_z \zetam\rangle \approx -\vao \sum_{\bm{k}_\perp,k_z>0}    k_z\frac{8\Theta}{(1-4\Theta^2)^2} \frac{|\zetap_{0,\bm{k}}|^2}{4} \nonumber \\ &\approx- \frac{\dot{a}}{4}\sum_{\bm{k}_\perp,k_z>0}  |\zetap_{0,\bm{k}}|^2 \approx -\frac{\dot{a}}{2}\sum_{\bm{k}}|\tilde{A}_{0,\bm{k}}|^2 \nonumber \\ &= -\dot{a}\anas(t=0),
\end{align}
where in the final steps we define the initial $\az $ as $\tilde{A}_{0,\bm{k}} $ and use  $\tilde{A}_{0,\bm{k}} \approx  \zetap_{0,\bm{k}}/2$. As expected, we have found that the 
$\vao\langle \zetap \partial_z \zetam\rangle/2$ term is exactly what is needed to cancel the expansion-induced growth term, $\dot{a}\anas$ in Eq.~\eqref{Eq: anastrophy conservation}, such that $\anas$ does not change  in time (averaged over the wave periods).
While not at all surprising, the calculation demonstrates the apparent ``fine tuning'' of the
linear solution when viewed from this perspective, highlighting how its disruption will necessarily decrease $|\langle \zetap \partial_z \zetam\rangle|$ and therefore drive wave-action anastrophy growth.

\section{Adaptive viscosity implementation} \label{app: adaptive viscosity}

The range of energies and scales involved in our simulations cover many orders of magnitude, while also 
differing significantly between $\bzp$ and $\bzm$ in the imbalanced phase. This poses a challenge for 
choosing the (hyper)-viscous dissipation coefficients $\nu^{\pm}$ to dissipate $\bzpm$ at small scales, because the nonlinear 
times, which  balance the dissipation times to set the dissipation scale of the turbulence, change significantly over the
course of the simulation (and differ between $\bzp$ and $\bzm$). Thus, rather than attempting to 
choose a functional form for $\nu^{\pm}$, which would require knowing \textit{a priori} the solution, 
we instead choose the co-moving dissipation scales, $\tilde{k}^{\rm diss}_{\perp}$ and $ k^{\rm diss}_{z}$ in the perpendicular and parallel directions, respectively, 
and adjust the dissipation coefficients $\nu^{\pm}_{\perp}$ and $\nu^{\pm}_{z}$ based on the local nonlinear time.

The idea is that the plus and minus energy fluxes arriving at $\tilde{k}^{\rm diss}_{\perp}$ and $ k^{\rm diss}_{z}$ are dissipated in one time-step $\delta t^{\pm}$ \cite{Borue97}:
\begin{eqnarray}
\dfrac{\tilde{\mathcal{\mathcal{E}}}^{\pm}(\tilde{k}^{\rm diss}_{\perp})}{\delta t^{\pm}}\sim \nu^{\pm}_{\perp}(\tilde{k}^{\rm diss}_{\perp}/a)^6\tilde{\mathcal{E}}^{\pm}(\tilde{k}^{\rm diss}_{\perp}),\\
\dfrac{\tilde{\mathcal{E}}^{\pm}(k^{\rm diss}_{z})}{\delta t^{\pm}}\sim \nu^{\pm}_{z}(k^{\rm diss}_{z}\va/\vao)^6\tilde{\mathcal{E}}^{\pm}(k^{\rm diss}_{z}),\label{Eq:Hyper-visc-z}
\end{eqnarray}
where  $\delta t^{\pm}$ is fixed by the maximum value of $
\vert\bzmp \vert $ by the standard Courant stability condition,
\begin{equation}
   \delta t^{\pm}=\dfrac{\rm{CFL}}{a^{-3/2}\pi n_{\perp} {\rm max} \vert \bzmp \vert /\tilde{L}_{\perp}}
\end{equation}
(here ${\rm CFL}$ is the standard Courant coefficient). 
 We choose, $\tilde{k}^{\rm diss}_{\perp} = 3/4(\pi n_{\perp}/\tilde{L}_{\perp})$, $k^{\rm diss}_{z} = 3/4(\pi n_{z}/L_{z0})$ and the coefficient ${\rm CFL}=1$.

\bibliographystyle{natbib}
\bibliography{rdtbib}
\end{document}